\pdfoutput=1
\documentclass[a4paper,twocolumn,epjc3]{svjour3}
\journalname{Eur. Phys. J. C}
\smartqed 
\makeatletter \let\cl@chapter\relax \makeatother

\usepackage{mathptmx}  
\usepackage[latin1]{inputenc}
\usepackage[T1]{fontenc}
\usepackage[english]{babel}
\usepackage{amsmath,amssymb}
\usepackage[usenames]{color}
\usepackage{graphicx,array,booktabs}
\usepackage[numbers,sort&compress]{natbib}
\usepackage[normalem]{ulem}
\usepackage[pdftex,hidelinks]{hyperref}
\hypersetup{colorlinks=true,
            linkcolor=blue,
            citecolor=red,
            filecolor=black,
            urlcolor=blue}
\hypersetup{pdfauthor={Mariano Cadoni, Edgardo Franzin and Matteo Tuveri},
            pdftitle={Hysteresis in eta/s for QFTs dual to spherical black holes}}
\usepackage[capitalise]{cleveref}
\crefformat{plural}{#2Eqs.~(#1)#3} 
\crefname{section}{Sect.}{Sects.}

\makeatletter\g@addto@macro\bfseries{\boldmath}\makeatother%

\def\be#1\ee{\begin{align}#1\end{align}}

\DeclareSymbolFont{lettersA}{U}{txmia}{m}{it}
	\DeclareMathSymbol{\R}{\mathord}{lettersA}{"92}
	\DeclareMathSymbol{\C}{\mathord}{lettersA}{"83}

\newcommand{\0}{\nonumber}
\newcommand{\ie}{i.e.}
\newcommand{\eg}{e.g.}
\def\O{\mathcal{O}}
\def\AdS{\text{AdS}}
\renewcommand{\geq}{\geqslant}
\renewcommand{\leq}{\leqslant}
\renewcommand{\ge}{\geqslant}
\renewcommand{\le}{\leqslant}
\def\a{\alpha}
\def\l{\lambda}
\def\s{\sigma}
\def\r{\rho}
\def\g{\gamma}
\DeclareMathOperator\Ima{Im}

\begin{document}

\title{Hysteresis in $\eta/s$ for QFTs dual to spherical black holes}

\author{Mariano Cadoni\thanksref{addr1,e1} \and Edgardo Franzin\thanksref{addr1,e2}\and and Matteo Tuveri\thanksref{addr1,e3}}

\thankstext{e1}{\href{mailto:mariano.cadoni@ca.infn.it}{mariano.cadoni@ca.infn.it}}
\thankstext{e2}{\href{mailto:edgardo.franzin@ca.infn.it}{edgardo.franzin@ca.infn.it}}
\thankstext{e3}{\href{mailto:matteo.tuveri@ca.infn.it}{matteo.tuveri@ca.infn.it}}

\authorrunning{M.~Cadoni, E.~Franzin and M.~Tuveri}

\institute{Dipartimento di Fisica, Universit\`a di Cagliari, \& INFN, Sezione di Cagliari, Cittadella Universitaria, 09042 Monserrato, Italy\label{addr1}}

\date{}

\maketitle%
\emergencystretch=1em

\begin{abstract}
We define and compute the (analogue) shear viscosity to entropy density ratio
$\tilde\eta/s$ for the QFTs dual to spherical AdS black holes both in Einstein and
Gauss-Bonnet gravity in five spacetime dimensions.
Although in this case, owing to the lack of translational symmetry of the background,
$\tilde\eta$ does not have the usual hydrodynamic meaning, it can be still interpreted as
the rate of entropy production due to a strain.
At large and small temperatures, it is found that $\tilde\eta/s$ is a monotonic increasing
function of the temperature. In particular, at large temperatures it approaches a constant
value, whereas, at small temperatures, when the black hole has a regular, stable extremal
limit, $\tilde\eta/s$ goes to zero with scaling law behaviour.
Whenever the phase diagram of the black hole has a Van der Waals-like behaviour,
\ie\ it is characterised by the presence of two stable states (small and large black holes)
connected by a meta-stable region (intermediate black holes), the system evolution must
occur through the meta-stable region and temperature-dependent hysteresis of $\tilde\eta/s$
is generated by non-equilibrium thermodynamics.
\end{abstract}

\section{Introduction}

In recent times, many efforts have been devoted to the investigation of the low-frequency,
hydrodynamic limit of quantum field theories~(QFTs) with holographic gravitational duals in
the AdS/CFT framework. This hydrodynamic limit is a powerful tool to compute transport
coefficients for strongly coupled QFTs, \eg\ the quark-gluon plasma phase of QCD\@.
In the hydrodynamic regime of thermal QFTs with gravitational duals, the shear
viscosity to entropy density ratio $\eta/s$ is of particular interest. In fact,
this ratio takes the \emph{universal} value $1/4\pi$ for all QFTs with Einstein gravity
duals~\cite{Policastro:2001yc,Buchel:2004qq,Benincasa:2006fu,Kats:2007mq,Landsteiner:2007bd,%
Iqbal:2008by,Buchbinder:2008nf,Edalati:2009bi}.
This has led to conjecture the existence of a fundamental lower bound $\eta/s\ge 1/4\pi$ ---
the Kovtun-Son-Starinets~(KSS) bound~\cite{Kovtun:2003wp} --- which is
supported both by energy-time uncertainty principle arguments and by
quark-gluon plasma experimental data~\cite{Kovtun:2003wp,Kovtun:2004de,Song:2010mg}.

In the usual setting of the AdS/CFT correspondence, the holographic dualities are utilised
to learn about transport coefficients in the hydrodynamic limit of strongly
coupled QFTs by investigating bulk gravity configurations, typically black branes.
However, this paradigm can be reversed and the properties of the dual QFT can be used
to infer about the behaviour of bulk gravity solutions.%
\footnote{For instance, this approach has been particularly
fruitful for the computation of the microscopic entropy of black holes.
In several cases, the Bekenstein-Hawking entropy has been matched
by counting states in the dual CFT --- see \eg\ Refs.~\cite{Strominger:1996sh,Cadoni:1998sg}.}
In this perspective, transport coefficients computed in the hydrodynamic limit
of the dual QFT can lead to a deeper understanding of black hole (BH) physics.
In particular, the aim of this paper is to better understand the rich thermodynamical phase
structure of AdS BHs with spherical horizons (characterised by meta-stabilities and
Van der Walls-like behaviour~\cite{Chamblin:1999tk,Chamblin:1999hg,Cvetic:2001bk}) by
investigating the relationship between the shear viscosity of the dual QFT and the
thermodynamics of these BHs.

It is well-known that the KSS bound can be violated by two different kinds
of effects: higher-curvature terms in the Einstein-Hilbert 
action~\cite{Brigante:2007nu,Brigante:2008gz,Ge:2009ac,Ge:2009eh,Cai:2009zv,Camanho:2010ru,%
Cremonini:2011iq,Jacobson:2011dz,Bhattacharyya:2014wfa,Sadeghi:2015vaa,Wang:2016vmm,Cadoni:2016hhd} 
and breaking of the translational or rotational symmetry of the
black brane background~\cite{Erdmenger:2010xm,Erdmenger:2011tj,Rebhan:2011vd,Mamo:2012sy,Davison:2015taa,%
Hartnoll:2016tri,Burikham:2016roo,Alberte:2016xja,Cadoni:2016hhd,Liu:2016njg}.
These two effects were the motivation to consider spherical AdS BHs, for which the translational
symmetry is intrinsically broken, both in general relativity (GR) and in a higher-curvature
theory, namely Gauss-Bonnet (GB) gravity.

The violation of the KSS bound in higher-curvature gravity theories, although not completely
understood, can be traced back to finite-$\mathcal{N}$, finite-$\l_{tH}$ effects and to the
inequality of the two central charges of the dual QFT~\cite{Buchel:2004di,Buchel:2008vz}.
This lends support to the possibility of formulating modified bounds on $\eta/s$, based for
instance on causality and positivity of energy in the dual QFT~\cite{Brigante:2008gz,%
Buchel:2009tt,Hofman:2009ug}.

On the other hand, the violation of the KSS bound due to the breaking of the translational symmetry has
a more fundamental nature. In this case, the shear viscosity does not have the usual hydrodynamic meaning
but might be interpreted as the rate of entropy production due to a strain~\cite{Hartnoll:2016tri,%
Davison:2015taa,Rebhan:2011vd,Mamo:2012sy,Burikham:2016roo,Romatschke:2009kr}.
In this framework, the behaviour of $\eta/s$ as a function of the
temperature $T$ is non-trivial~\cite{Cremonini:2012ny,Cremonini:2011ej} and carries information about the infrared~(IR)
and ultraviolet~(UV) behaviour of the QFT, the existence of global diffusive modes of the system and
the nature of the effect responsible for the breaking of translational invariance.
For instance, when this breaking is generated by the presence of a non-homogeneous scalar field
in the bulk, the behaviour of $\eta/s$ at small $T$ is determined by the flow of the QFT in the IR\@.
If the translational invariance is restored in the IR then $\eta/s$ goes to a constant as $T\to0$,
signalising the presence of an IR collective diffusive mode. Conversely, if the translational
invariance is not restored, $\eta/s$ scales as $T^{2\nu}$ for $T\to0$ and
the IR geometry in $D+2$ dimensions is typically $\AdS_2\times\R_D$~\cite{Hartnoll:2016tri}.
We will discuss the general validity of this behaviour for spherical BH backgrounds.
In this case, the translational symmetry is intrinsically broken and cannot be restored in the IR\@,
but holds only in the UV, where the spherical horizon can be approximated by a plane.

In a recent letter~\cite{Cadoni:2017fnd}, a definition of the shear viscosity
for QFTs living in manifolds whose spatial sections are spheres is proposed. This has allowed us to compute
the spherical analogue of the shear viscosity $\tilde\eta$ for QFTs dual to five-dimensional
AdS-Reissner-Nordstr\"om (AdS-RN) BHs in GR\@. Here, a detailed derivation of these results
is presented and the discussion to five-dimensional asymptotically AdS neutral and charged BHs in GB gravity is extended.

We start by defining the (analogue) shear viscosity $\tilde\eta$ for a QFT living
on a $D$-sphere in the hydrodynamic limit.
Although the shear viscosity does not have the usual interpretation
pertaining to a QFT in a translation-invariant background,
it is shown that it still satisfies a Kubo formula and can be interpreted
in terms of entropy production due to a strain.

For a given tensorial perturbation of the spherical background it is possible to define three different correlators 
corresponding to shear, sound and transverse propagating modes~\cite{Brigante:2007nu}. 
If all the background symmetries are unbroken they lead to the same value of 
the shear viscosity. If this is not the case, in general the value 
of $\tilde \eta$ becomes channel dependent~\cite{Ciobanu:2017fef}. 
Therefore, in the case under consideration there will be three
different determinations of the viscosity for shear, sound and transverse modes.
In this paper the focus will be only on the shear viscosity for transverse perturbations, 
for  which  computations are  easier.

Following the approach of Refs.~\cite{Hartnoll:2016tri,Lucas:2015vna},
we compute $\tilde\eta/s$ for QFTs dual to five-dimensional
asymptotically AdS neutral and charged BHs in GR and GB gravity.
By considering linear perturbations of the field equations, the computation of $\tilde\eta$ is reduced
to the determination of the non-normalisable mode of the perturbation
evaluated at the horizon. The perturbation satisfies a linear second-order differential
equation analogous to a massive scalar equation in a curved background whose non-vanishing
mass term encodes the breaking of the translational invariance.
Whereas the large and small $T$ behaviours of $\tilde\eta/s$ are determined analytically, its global
behaviour is determined numerically.
It is shown that for certain regimes of the temperature $\tilde\eta/s$ is a monotonic function.
It saturates the KSS bound at large~$T$ (or the GB coupling constant dependent bound in the GB
case), where the translational invariance is restored. When the BH has a regular, stable extremal limit
$\tilde\eta/s$ goes to zero with a $T^{2\nu}$ scaling law at small temperatures.

An interesting and somehow unexpected behaviour of $\tilde\eta/s$ emerges in the parameter
regions where the BH has a Van der Walls-like behaviour, 
characterised by the presence of both a second and a first order phase transition.
Once the control parameter (the GB coupling constant or the BH charge) falls 
below a critical value, the system undergoes a second order phase transition. 
In this condition, BHs may undergo a first-order phase transition from small to large BHs 
controlled by the temperature. Small and large black 
holes are connected through a meta-stable intermediate region.
As a result, it is found that $\tilde\eta/s$ exhibits a temperature-dependent hysteresis 
and, close to the phase transition, it becomes multi-valued as expected for a 
first-order phase transition~\cite{Erdmenger:2010xm}.
We further explain this behaviour in terms of the non-equilibrium thermodynamics underlying the
Van der Walls-like phase portrait.

The structure of the paper is as follows.
In \cref{sec:hydrodynamics} we discuss the problems related to the hydrodynamic
limit for QFTs dual to spherical BHs, the definition and the computation of $\tilde\eta$
and we derive the Kubo formula for $\tilde\eta$.
In \cref{Sec:BHsol} we review known facts about solutions, thermodynamics, phase structure and
perturbations for AdS spherical BHs in GR and GB gravity.
In \cref{sec:etatos} we give the general formula for $\tilde\eta/s$, compute its
large and small $T$ behaviours, give the numerical results for its global behaviour
and discuss its relationship with the thermodynamical phase portrait of the dual BH solutions.
Finally, in \cref{sec:conclusions} we state our conclusions.

Throughout this paper indices $a, b, \dots$ refer either to the whole $D+2$-dimensional bulk spacetime
or to its $D+1$-dimensional conformal boundary, while $i, j, \dots$
refer to the transverse $D$-dimensional spatial sections.

\section{Hydrodynamic limit for QFTs dual to spherical black holes and the analogue viscosity\label{sec:hydrodynamics}}

Relativistic hydrodynamics is an effective long-distance description for a
classical or quantum many-body system at non-zero temperature. In particular,
it can be used to describe the non-equilibrium real-time macroscopic slow evolution of the system,
both in space and time, with respect to a certain microscopic scale.

In the holographic framework of the AdS/CFT correspondence, the QFT lives in the
boundary of a certain gravitational bulk region. In some cases, the QFT can be
described by a kinetic theory and the microscopic scale is determined by the mean
free path of particles~$l_\text{mfp}$ and the typical momentum scale of the
process~$k$. When the kinetic theory is absent or unknown, it is still possible
to give a thermal description and interpret the inverse temperature as the
microscopic scale~\cite{Baier:2007ix,Kovtun:2012rj}.
Thus, the hydrodynamic limit of a QFT corresponds to large relaxation time,
\ie\ small frequencies, and large scales compared to the typical
one of the system, \ie\ $\tilde{\lambda}\gg 1/T\sim l_\text{mfp}$, where
$\tilde\lambda$ is the wavelength of the excitations of the system.

In general, the existence of a hydrodynamic description is essentially due to the
presence of conserved quantities, \ie\ to the isometries of the system, whose
densities can evolve (oscillate or relax to equilibrium) at arbitrarily long times
provided the fluctuations are of large spatial size. Correspondingly, the expectation
values of such densities are the hydrodynamic fields. However, it is still possible
to give a hydrodynamic description of a system \emph{without} conserved quantities
in terms of expansion in derivatives of hydrodynamic fields (as the fluid
velocity)~\cite{Baier:2007ix}.
This approach is followed to formulate the hydrodynamic description of a fluid
in a spherical background holographically dual to AdS spherical BHs.

On the sphere, due to its intrinsic geometry, the translational invariance is broken.
As a consequence, the momentum is not conserved and it is not possible to define an associated
conserved current. At first sight, this should prevent us from studying transport
coefficients as the shear viscosity~$\eta$ which is, by definition, a measure of the
momentum diffusivity due to a strain in a fluid.
Hence, in principle, without translational symmetry it is not possible to define
a conserved current, from which the Fick law of diffusion~\cite{Son:2007vk} can be derived.
Nevertheless, as seen below, these difficulties can be circumvented and 
a rigorous definition of $\eta$ for the hydrodynamic limit of a QFT in a spatial
background without translational isometries can be given.

Consider a QFT living on the boundary of $\AdS_{D+2}$ whose spatial sections
have spherical topology. Although bulk BHs allow for dual
QFTs living on a sphere~\cite{Gubser:1998bc,Witten:1998qj,Cho:2002hq,Neupane:2009zz},
the explicit form of the holographically dual QFT is not of interest here.
However, its hydrodynamic limit can be studied in the sense described above.
The boundary metric is conformal to $\R\times S^D$
\be\label{boundary}
ds^2=\frac{r^2}{L^2}\left(-dt^2 + L^2\,d\bar{\Omega}_D^2\right),
\ee%
where $d\bar{\Omega}_D^2=\bar{g}_{ij}\,dx^i dx^j$ is the metric of a $D$-sphere. In this case, due to the spherical shape of
the boundary, the metric perturbations used to describe the non-equilibrium real-time
macroscopic slow evolution of the system are characterised by two parameters,
the relaxation time or the frequency $\omega$ and $L/\ell$ which ``measures'' angular distances on the sphere.
The integer number $\ell$ parametrises the eigenvalue of the Lichnerowicz operator on the sphere
(see \cref{GreenFunction} below) and is analogous to the momentum scale $k$ for a flat topology.
In the spacetime~\eqref{boundary}, the hydrodynamic limit of the holographic QFT is defined
as the limit in which the metric perturbations have slow relaxation time and are much larger
than the typical scale of the system, \ie\ $\omega\to0$ and $L/\ell\gg 1/T$.
Since we are dealing with a $D$-sphere, the number
$\ell$ cannot be arbitrarily small, \ie\ there is a minimum value
$\ell_0$~\cite{Kodama:2003jz,Ishibashi:2003ap,Gibbons:2002pq} which corresponds to a maximum
spatial scale, and to a maximum size for the global modes propagating on the sphere.
On the contrary, in flat space, there is no constraint on the values of $k$.
Thus, one can set $k\to0$ which corresponds to fluctuations of very
large (in principle infinite) wavelength.

\subsection{Hydrodynamics in curved spacetime}

Relativistic hydrodynamics for a fluid in curved spacetimes can be formulated
starting from the following definition for the stress-energy tensor~\cite{Romatschke:2009kr,Baier:2007ix}
\be\label{TabBordo}
T^{ab}=\epsilon u^a u^b + T_{\perp}^{ab}\,,
\ee%
where $\epsilon$ is the energy density and the fluid velocity $u^a$ (commonly evaluated
in the frame in which the fluid is at rest) is timelike.
The tensor $T_{\perp}^{ab}$ is the spatial part of the stress-energy tensor and
it is made by time-independent functions of the hydrodynamic variables $\epsilon$,
$u^a$ and their derivatives.
In a generic curved background, it is not always possible to define globally
conserved currents associated with symmetries of the system. However, the hydrodynamic
equations can always be derived by requiring the stress-energy tensor
to be covariantly conserved, \ie\ $\nabla_a T^{ab}=0$.
In general, the hydrodynamic modes are infinitely slower than all other
modes and the latter can be integrated out. Thus, all quantities appearing in
the hydrodynamic equations are averaged over these fast modes and are functions
of the slow-varying hydrodynamic variables.

\Cref{TabBordo} can be expanded in powers of derivatives of the velocity,
and at first order, the most general expansion is given by
\be\label{dissipativeT}
T^{ab}= \left(\epsilon + P\right) u^a u^b + P g^{ab} + \Pi^{ab}\,,
\ee%
where $P=P(\epsilon)$ is a scalar function and it can be interpreted as the
thermodynamical pressure. The tensor $\Pi^{ab}$ contains the derivatives of the
fluid velocities, \ie\ the dissipative contributions to~$T^{ab}$. Its explicit
form is given by~\cite{Romatschke:2009kr,Baier:2007ix}%
\footnote{For a rank-2 tensor,
${}^{\langle}A^{ab}{}^{\rangle}=A^{\langle{}ab\rangle}\equiv
\frac{1}{2}\Delta^{ca}\Delta^{db}\left(A_{ab}+A_{ba}\right)
-\frac{1}{D}\Delta^{ab}\Delta^{cd}A_{cd}$,
where $\Delta^{ab}$ is a symmetric and transverse tensor given by
$\Delta^{ab} = g^{ab} + u^a u^b$. In the local rest frame,
it is the projector tensor on the spatial subspace.}
\be%
\Pi^{ab} = &-\eta\sigma^{ab}
- \eta\tau_{\Pi}\left[{}^{\langle}\mathcal{D}\sigma^{ab\rangle}
+ \frac{1}{D}\sigma^{ab}(\nabla_cu^c)\right]\0\\
&+\kappa \left[R^{\langle{}ab\rangle}
- (D-1) u_c R^{c\langle{}ab\rangle{}d}u_d\right] + \cdots\label{dissipationTerms}
\ee%
where the dots represent the non-linear terms in the fluid velocity and $\eta,\tau_{\Pi}$, $\kappa$ are transport 
coefficients.
The symbol $\mathcal{D}$ represents the derivative with respect to the velocity
direction, \ie\ $\mathcal{D}=u_a\nabla^a$. The tensor $\sigma^{ab}$ is a symmetric,
transverse $u_a\sigma^{ab}=0$ and traceless $g_{ab}\sigma^{ab}=0$ tensor
constructed with the first derivative in the fluid velocity given by
$\sigma^{ab}=2\vphantom{u}^{\langle}\nabla^{a}u^{b\rangle}$.
The parameter $\eta=\eta(\epsilon)$ is the shear viscosity and $\tau_{\Pi}$
is the relaxation time.
%
%
%
%

\subsection{Kubo formulas and the analogue viscosity}

The Kubo formula relates thermal correlators to kinetic coefficients such as
dissipative ones. For a relativistic QFT in flat spacetime, the Kubo formula
gives a general definition of the shear viscosity in terms of the retarded Green
function for the stress-energy tensor~\cite{Kovtun:2004de,Policastro:2002se,Hartnoll:2016tri}
\be\label{shearKubo}
\eta=\lim_{\omega\to0}\frac{1}{\omega}\Ima{} G_{T^{ij}T^{ij}}^R(\omega,k\to0)\,,
\ee%
where $i=x,\ j=y$, $T^{ij}$ are the spatial components of the stress-energy
tensor, $\omega$ and $k$ are the frequency and wave vector of the perturbation.
When the translational invariance is
preserved and a hydrodynamic limit exists, \cref{shearKubo} becomes the Kubo
formula for the transverse momentum. In this case, $\eta$ defined by
\cref{shearKubo} coincides with the usual hydrodynamical definition in terms of
conserved quantities obtained from the Einstein relation $C=\eta/sT$, where $C$
is the diffusion constant appearing in the Fick law~\cite{Hartnoll:2016tri}.
In a holographic setup based on the AdS/CFT correspondence,
$G_{T^{ij}T^{ij}}$ can be calculated using the usual AdS/CFT rules by considering
small perturbations of the bulk metric.

In order to extend the Kubo formula~\eqref{shearKubo} to spherical backgrounds,
small metric perturbations around the boundary background
metric~\eqref{boundary} are considerd, \ie\ $g_{ab}\to g_{ab} + h_{ab}$.
In general, three different types of perturbations can be considered:
shear, sound and transverse (scalar) modes. The behaviour of these modes 
will be encoded in three different correlators $G_{1,2,3}(\omega,k)$. 
In the translationally invariant case (and also when translation invariance 
is broken by external matter fields) at $k=0$ these three functions are equal, 
owing to rotational symmetry~\cite{Brigante:2007nu}. By contrast, in the 
spherical case under consideration $k$ cannot be taken to zero by construction 
and the correlators will be different. Thus, in general, any definition of 
the shear viscosity in a spherical background based on linear response to 
small disturbances will be channel-dependent. In this paper the focus will be on 
the transverse perturbations. The computations for the sound and shear 
channel is left for future investigations.

Choosing transverse and traceless perturbations with
$h_{ab}=0$ if $(a,b)\neq(i,j)$, $h_{ij}=h_{ij}(t,\mathbf{x})$ (where $\mathbf{x}$ denote
the angular directions) in \cref{dissipativeT}
and considering the fluid at rest, \ie\ $u^a= (1,\mathbf{0})$, we obtain
\be%
T^{ij} = -Ph_{ij}-\eta\dot{h}_{ij}+\eta\tau_{\Pi}\ddot{h}_{ij}
-\frac{\kappa}{2}\left[(D-2)\ddot{h}_{ij}+L^2\bar{\bigtriangleup}_Lh_{ij}\right],\label{shearmodes}
\ee%
where $\bar{\bigtriangleup}_L=\bar{\nabla}_k\bar{\nabla}^k$ is the Lichnerowicz
operator and it corresponds to a generalisation of the Laplacian for the $D$-sphere,
with $D\geq3$. \Cref{shearmodes} is analogous to the one obtained in
Ref.~\cite{Baier:2007ix} for planar topology.
As required by linear response theory, the retarded Green function
for the tensor channel is computed by choosing a harmonic time dependence for the perturbation,
$h_{ij}(t,\mathbf{x})=e^{-i\omega t}\,h_{ij}(\mathbf{x})$ and by expanding in
hyper-spherical harmonics~\cite{Rubin:1983be,Rubin:1984tc,Higuchi:1986wu},
the retarded Green function can be extracted from \cref{shearmodes},
\be%
G_{T^{ij}T^{ij}}^R(\omega,\ell)&=-P-i\omega\eta-\omega^2\eta\tau_{\Pi}\0\\
&\phantom{=}-\frac{\kappa}{2}\left[(D-2)\omega^2+L^2\gamma\right],\label{GreenFunction}
\ee%
where $\gamma=\ell(\ell+D-1)-2$ are the eigenvalues of the Lichnerowicz operator and
$\ell=1,2,3,\ldots$ is an integer associated with the hyper-spherical harmonic
expansion. The eigenvalues $\g$ are positive and form a discrete
set~\cite{Higuchi:1986wu,Kodama:2003jz,Ishibashi:2003ap,Gibbons:2002pq}.
Given the retarded Green
function above, the dissipative coefficients $\eta$ and $\tau_{\Pi}$ can be extracted.
In particular, the analogue of shear viscosity in the hydrodynamic limit for a QFT in
a spatial spherical background in the transverse channel can be defined as,
\be\label{analogueKubo}
\tilde{\eta}\equiv-\lim_{\omega\to0}\frac{1}{\omega}\Ima G_{T^{ij}T^{ij}}^R(\omega,\ell\to\ell_0)\,,
\ee%
where $\ell_0$ is the minimum value of $\ell$.
Notice that the shear viscosity $\tilde\eta$ in \cref{analogueKubo} is defined
as the $\ell\to\ell_0$ limit of the retarded Green function in analogy with \cref{shearKubo}.
In planar hydrodynamics, the $k\to0$ limit describes long wavelength modes and
probes large scales on the plane.
In the spherical case, the $\ell\to\ell_0$ modes probe large angles on the sphere.

It is also important to stress that, with respect to the planar case,
the expression in square brackets in \cref{shearmodes} has an additional contribution
to the stress-energy tensor ruled by the transport coefficient $\kappa$. However this contribution drops out in the 
Kubo formula~\eqref{analogueKubo}, when we take the imaginary part of the Green function.

We conclude with some remarks about the conservation of the stress-energy tensor.
For translation-invariant backgrounds, the conservation of the stress-energy tensor leads to the
conservation of global currents and to the Fick law~\cite{Son:2007vk}.
More generally, from the projection of
$\nabla_a T^{ab}$ along the fluid velocity $u_b$, second-order
hydrodynamics can be related with the second law of thermodynamics~\cite{Baier:2007ix}. In particular,
by using \cref{dissipativeT,dissipationTerms} at linear order the following can be found
\be\label{entropyproduction}
\dot{s}=\frac{\eta}{2T}\,\sigma_{ij}\sigma^{ij}\,,
\ee%
where $s$ is the entropy density.
\Cref{entropyproduction} represents the rate of entropy production in a fluid
due to a slowly varying strain $h_{ij}$ and it can be used to define 
the shear viscosity~\cite{Hartnoll:2016tri}.

In this case, only local conservation can be considered with the background metric~\eqref{boundary},
since the translational invariance is broken and the Fick law
is not satisfied but \cref{entropyproduction} still holds.

\section{Black-hole solutions in five dimensions\label{Sec:BHsol}}

The field equations of five-dimensional Einstein-Gauss-Bonnet gravity sourced by
any form of matter fields described by the stress-energy tensor $T_{(M)}{}_a^b$
are~\cite{Myers:1988ze,Cai:2001dz,Cai:2003kt}
\be\label{EGBEOM}
G_{(1)}{}_a^b + \a_2\,G_{(2)}{}_a^b = 8\pi G_5\,T_{(M)}{}_a^b\,,
\ee%
where $G_{(1)}{}_a^b\equiv R_a^b-\frac{1}{2}R\delta_a^b$ is the Einstein tensor,
$\a_2$ is the GB coupling constant, $G_5$ is the five-dimensional Newton constant,
and $G_{(2)}{}_a^b$ is the GB contribution,
\be%
G_{(2)}{}_a^b \equiv R_{ca}{}^{de} R_{de}{}^{cb} - 2R_d^c R_{ca}{}^{db} - 2R_a^c R_c^b + R R_a^b\0\\
-\frac{1}{4}\delta_a^b\left(R_{cd}{}^{ef}R_{ef}{}^{cd} - 4R_c^d R_d^c + R^2\right).\label{GBtensor}
\ee%

For later convenience we define $\l\equiv\a_2/L^2$, being $L$ the AdS length.
Throughout this paper, the source term contains only a negative cosmological constant
and an electromagnetic field.
In particular, we consider static BH solutions to~\eqref{EGBEOM} \ie\ solutions with spherical horizons in the form
\be\label{backgroundmetric}
ds^2=-f(r)\,dt^2+\frac{dr^2}{f(r)}+r^2\,d\bar{\Omega}_{3}^2\,.
\ee%
For the AdS-RN BHs of GR the metric function is
\be\label{eq:fRN}
f_\text{RN}(r) = 1+\frac{r^2}{L^2}- \frac{8 G_5 M}{3\pi r^2}+\frac{4\pi G_5 Q^2}{3r^4}\,,
\ee%
while, in the branch that allows for BH solutions,
the metric function for GB gravity is
\be\label{bhmetric}
f_\text{GB}(r)= &1+\frac{r^2}{2\l L^2}\Bigg[1-\0\\
&\sqrt{1-4\l L^2\left(\frac{1}{L^2}-\frac{8 G_5 M}{3\pi r^4}+\frac{4\pi G_5Q^2}{3 r^6}\right)}\Bigg].
\ee%
In \cref{eq:fRN,bhmetric}, $M$ and $Q$ are, respectively, the BH mass and charge.

\subsection{Black holes in Gauss-Bonnet gravity\label{section:chargedGB}}

As in the black brane case, asymptotically AdS BH solutions of GB gravity exist only for $\l<1/4$.
Moreover, it is well-known that the unitarity bounds for the dual QFT
constrain the value of $\l$~\cite{Brigante:2007nu, Ge:2008ni, Ge:2009eh},
so that in this paper we will take $\l$ in the following range $0<\l\le9/100$.

The BH horizons are determined by the positive zeroes of the function
\be\label{horizonseq}
h(Y)=\frac{Y^3}{L^2}+ Y^2-\s Y+\r\,,
\ee%
where $Y=r^2,\, \s= 8 G_5 M/3\pi -\l L^2 ,\, \r = 4\pi G_5 Q^2/3$.
The BH becomes extremal when $h'(Y)=0$.

Asymptotically AdS BH solutions with inner ($r_-$) and outer ($r_+$) horizons exist for
\be\label{massbound}
M\ge\frac{3\pi}{8 G_5} \left[\l L^2+ \frac{L^2}{3}\left(z_0^2+2z_0\right)\right],
\ee%
where $z_0$ is the real, positive solution of the cubic equation
$2z^3 + 3z^2 -27\r/L^4=0$. When the inequality is saturated, the inner and outer
horizons merge, \ie\ the BH becomes extremal and in the near-horizon regime the
solution factorises as $\AdS_2\times S^3$
\be\label{metrichorizonextreme}
ds^2=- \frac{r^2}{l^2}\,dt^2+\frac{l^2 dr^2}{r^2}+r_0^2\,d\bar{\Omega}_3^2\,,
\ee%
where $r_0$ is BH radius at extremality, determined by the solution $Y_0=r_0^2$
of the cubic equation
\be\label{horizonextreme}
h_\text{ext}(Y)=\frac{2Y^3}{L^2}+Y^2-\r=0\,,
\ee%
and $l$ is the $\AdS_2$ length
\be\label{ads2length}
l^{-2}= \frac{2 h''(r_0)}{r_0^2+ 2\l L^2} = \frac{2(6r_0^2+2L^2)}{L^2(r_0^2+ 2\l L^2)}\,.
\ee%

The BH thermodynamical parameters temperature $T$, mass $M$ and entropy $S$ can
be expressed in terms of the horizon radius $r_+$ as~\cite{Cai:2003kt}
\be%
T(r_+) &= \frac{\frac{4r_+^4}{L^2}+2r_+^2-\frac{8\pi G_5Q^2}{3r_+^2}}{4\pi r_+(r_+^2+2\l L^2)},\label{MTS:temperature}\\
M(r_+) &= \frac{3\pi r_+^4}{8 G_5}\left(\frac{1}{L^2}+\frac{1}{r_+^2}+\frac{\l L^2}{r_+^4}
+\frac{4\pi G_5Q^2}{3r_+^6}\right),\label{MTS:mass}\\
S(r_+) &= \frac{\pi^2r_+^3}{2G_5}\left(1+\frac{6\l L^2}{r_+^2}\right).\label{MTS:entropy}
\ee%

The spherical geometry of the horizon introduces another scale in the system,
\ie\ the radius of the sphere, which couples in a non-trivial way to the
higher-curvature terms in the equations of motion~\eqref{EGBEOM}.
This scale introduces dependence on the GB coupling in the mass bound~\eqref{massbound} and
in the thermodynamical expression \labelcref{MTS:mass,MTS:temperature,MTS:entropy}.
As a result, the thermodynamical and near-horizon behaviours of the GB BHs are
completely different from their brane counterparts. Indeed, for charged GB black
branes, such behaviours are universal, \ie\ do not depend on $\l$, and are
essentially the same of the RN black branes of GR~\cite{Cadoni:2016hhd}.

Notice that although the extremal radius $r_0$ is determined only by the BH charge
and the cosmological constant, the $\AdS_2$ length $l$ and hence the extremal
geometry~\eqref{metrichorizonextreme} depend on the GB coupling constant.
Notice also that the expression in the parenthesis in \cref{MTS:temperature} is
proportional to $h_\text{ext}(Y_+)$ meaning that the extremal geometry is obtained
at zero temperature.

The thermodynamical parameters~\labelcref{MTS:mass,MTS:temperature,MTS:entropy}
near-extremality are
\be%
T(r_+) &= \frac{2}{\pi L^2} \frac{3r_0^2+L^2}{r_0^2+2\l L^2} (r_+-r_0)
+\O\left({(r_+-r_0)}^2\right),\label{Text}\\
M(T) &= \frac{3\pi}{8 G_5}\left(\frac{3r_0^4}{L^2} + 2 r_0^2+\l L^2\right)\0\\
&\phantom{=}+ \frac{3\pi^3}{8 G_5}\frac{L^2 (r_0^2+2\l L^2)^2}{L^2 + 3r_0^2}\,T^2
+\O\left(T^3\right),\label{Mext}\\
S(T) &= \frac{\pi^2 r_0^3}{2G_5}\left(1+\frac{6\l L^2}{r_0^2}\right)\0\\
&\phantom{=}+ \frac{3\pi^3}{4G_5} \frac{L^2(r_0^2+2\l L^2)^2}{L^2+3r_0^2}\,T
+\O\left(T^2\right)\label{Sext}.
\ee%
The first terms in \cref{Mext,Sext} represent,
respectively, the BH mass and entropy at extremality.

\subsection{Phase structure of AdS-Reissner-Nordstr\"om black holes\label{subSec:AdSBHQ}}

Although the metric function $f_\text{GB}$ in \cref{bhmetric} is singular for $\l=0$,
the thermodynamical behaviour of the charged AdS-RN solution can be simply obtained
by putting $\l=0$ in \cref{massbound,MTS:mass,MTS:temperature,MTS:entropy}.

To characterise the phase structure of these BHs, a distinction between two cases can be made:
fixed electric potential or fixed electric charge~\cite{Chamblin:1999tk,Chamblin:1999hg}.
In this paper we only discuss the canonical ensemble, \ie\ we work at 
fixed charge. We will not consider the grand canonical ensemble, \ie\ the case of 
fixed chemical potential.
As the charge of BH decreases to a critical value $Q_c$, the system undergoes a
second-order phase transition. Below the critical charge, there are three possible
branches of solutions that depend on the radius and therefore on the temperature
of the system. For small temperatures, a small BH is the only locally stable solution;
as the temperature increases, a meta-stable configuration describing
intermediate BHs can be found; for sufficiently high temperatures, large BHs are globally preferred.
The evolution from small to large BHs through the meta-stable region corresponds
to a first-order phase transition.
Above the critical charge, the BH solution is always globally preferred.
This behaviour can be understood by analysing the temperature as a function of the
BH radius given by \cref{MTS:temperature} with $\l=0$. For $Q>Q_c$ it is a monotonic
function, whereas it develops local extrema for $0<Q<Q_c$ and an inflection point
for $Q=Q_c$. Notice that the case $Q=0$ is not included in the range of existence 
of the first order phase transition. In fact, $Q=0$ corresponds to the 
AdS-Schwarzschild BH.
The phase portrait of the AdS-RN BHs is very similar to a liquid/gas
Van der Waals phase transition where the BH temperature plays the role of the pressure,
the BH radius that of the volume and the BH charge that of the
temperature~\cite{Chamblin:1999tk,Chamblin:1999hg}.
This portrait has been extended by Kubiz\v{n}\'ak \emph{et al.} in
Refs.~\cite{Kubiznak:2012wp,Kubiznak:2016qmn} and to topological AdS BHs in
massive gravity~\cite{Hendi:2017fxp}.

We make a brief comment on the zero-charge limit.
For $Q=0$, the metric \eqref{eq:fRN} reduces to that of an AdS-Schwarzschild 
BH\@. However, from the thermodynamical  point of view, this 
limit is singular. There is a discontinuity at $Q=0$, and the phase diagram of
an AdS-Schwarzschild BH cannot be obtained as the $Q\to 0$ limit of the 
AdS-RN one. In fact, the BH temperature as a function of the radius,~\cref{MTS:temperature}, 
when $Q=0$ becomes a monotonic function and shows no Van der Waals-like behaviour 
as in the RN case~\cite{Cadoni:2017fnd}.

\subsection{Phase structure of neutral Gauss-Bonnet black holes\label{section:ngbbh}}

Neutral GB BH solutions and their thermodynamical parameters are obtained by
putting $Q=0$ in \cref{bhmetric,MTS:mass,MTS:temperature,MTS:entropy}.
These BHs are characterised by the absence of a regular, zero temperature extremal
limit which, in turn, means the absence of an IR fixed point for the dual QFT
in the holographic description. For positive $\l$, the $T=0$ extremal limit is a
state with $r_+=0$, zero entropy and positive non-vanishing mass. Therefore, the
small temperature thermodynamical behaviour is always singular.

Neutral GB BHs exhibit an interesting phase structure. Different from Einstein
gravity, where small BHs are not stable and a thermal AdS state is energetically
preferred~\cite{Hawking:1982dh,Birmingham:1998nr}, in GB gravity there exists a
stable small BH\@.%
\footnote{Small BHs can be gravitationally unstable for values of $\l$
larger than those considered here~\cite{Konoplya:2017ymp}.}
It starts with a small positive free energy, becomes unstable
and evolves to a thermal AdS phase. Additionally, we also have the usual stable
BH phase for large radii~\cite{Cho:2002hq}.
By inspecting the behaviour of the specific heat and the free energy, it is found
that the phase structure of neutral GB BHs strictly depends on the values of the
GB coupling constant and the BH radius~\cite{Cai:2001dz}. For values of the GB
coupling constant below the critical one, $\l_c=1/36$, there are three different
branches of solutions that correspond to small, intermediate and large BHs.
The specific heat is positive in the first and third branch, whereas it is negative
in the second branch. This behaviour is a consequence of the fact that $T(r_+)$
given by \cref{MTS:temperature} with $Q=0$ is monotonically increasing for
$\l>\l_c$, whereas it develops local extrema for $\l<\l_c$~\cite{Cai:2001dz}.
For $\l\geq\l_c$ the second branch disappears and BHs are always locally stable
but not necessarily globally preferred. Computing the free energy one finds that
the BH solution is globally stable and energetically preferred with respect to
thermal AdS in the parameter region $\l_1(r_+) \le\l\le\l_2(r_+)$, where $\l_1(r_+)$
and $\l_2(r_+)$ are some functions of the horizon radius~\cite{Cai:2001dz}.
Outside this region a Hawking-Page phase transition exists, BHs become globally
unstable and thermal AdS is energetically preferred.
Therefore, in the parameter region where BHs are energetically preferred with respect
to thermal AdS, the phase diagram of uncharged GB BHs has the same Van der Waals
form described in the previous section for AdS-RN BHs, with the GB parameter
$\l$ playing the role of the BH charge~$Q$. 

Analogously to the $Q\to 0$ limit, the limit $\l\to0$ is singular 
from the thermodynamical point of view. In fact for $\l\to0$ 
the phase diagram of an AdS-Schwarzschild BH cannot be recovered. First, the metric \eqref{bhmetric} 
becomes singular. Second, similarly to what we have seen for charged BHs 
in GR, the temperature as a function of the horizon radius exhibits a a discontinuous 
behaviour in the $\l\to0$ limit. The limits $Q\to 0$ and $\lambda \to0$ 
have a similar singular behaviour also in the case of charged GB BHs.

\subsection{Phase structure of charged Gauss-Bonnet black holes\label{section:chargedbh}}

The thermodynamical description of charged GB BHs is determined by the GB coupling
constant $\l$ and the charge $Q$.
There are critical values of these parameters such that these BHs can exhibit
the typical Van der Waals gas behaviour in the $T$-$S$ plane~\cite{Hu:2013cia,Zeng:2016aly}.%
\footnote{This is analogous to consider the cosmological constant as a pressure
term in the BH equation of state~\cite{Frassino:2014pha}.}
Thus, charged GB BHs possess both the Hawking-Page phase transition~\cite{Hawking:1982dh,Cvetic:2001bk}
and a second-order one~\cite{Zeng:2016aly}.
The former represents the transition from a stable AdS thermal state
to a stable BH spacetime.
When $T_c$ is the $r_+$-dependent critical value of the temperature and $r_c^2=6\l L^2$.
Then, for $T>T_c$ and $r_+>r_c$ (or $T<T_c$ and $r_+<r_c$)
AdS is preferred with respect to the BH, whereas for $T<T_c$ and $r_+>r_c$ (or $T>T_c$ and $r_+<r_c$),
the BH is preferred with respect to AdS.
It is remarkable that due to presence of $\l$ and $Q$, the standard critical point
becomes a critical line in the $T$-$r_+$ phase diagram~\cite{Cvetic:2001bk}.

Again, the phase portrait has the Van der Waals-like form described in
\cref{section:ngbbh,subSec:AdSBHQ} if one considers only the parameter region where
the BH phase is globally preferred with respect to the thermal AdS phase and if
one holds either $Q$ or $\l$ fixed. In the former (latter) case, at the critical
value $\l_c$ ($Q_c$) the system undergoes a second-order phase transition. For
$\l<\l_c$ ($Q<Q_c$), varying the temperature we have again a stable small BH phase
and a stable large BH phase connected by a meta-stable phase. Moreover, the function
$T(r_+)$ always has the typical behaviour described in \cref{subSec:AdSBHQ,section:ngbbh}.

\subsection{Linear perturbations in Einstein-Gauss-Bonnet gravity\label{linearpertub}}

In this section, we study linear tensorial perturbations about the background~\eqref{backgroundmetric}
in Einstein-Gauss-Bonnet gravity, \ie\ $g_{ab}\to g_{ab} + h_{ab}$.
After suitable manipulations, the linearised equations of motion~\eqref{EGBEOM} are
\be\label{perturbedGB}
\delta R_i^j + \l L^2\,\delta G_{(2)}{}_i^j
+ 8\pi\,G_5\left(T_{(M)}{}_i^k h_k^j - \frac{\delta T_{(M)}{}_{ij}}{h_{ij}}\,h_i^j\right)=0\,,
\ee%
where $\delta T_{(M)}{}_{ij}=\left(\frac{\delta T_{(M)}{}_{ij}}{h_{ij}}\right)h_{ij}$ and
the explicit form of the tensors
$\delta R_i^j$ and $\delta G_{(2)}{}_i^j$ can be found in Refs.~\cite{Dotti:2004sh,Dotti:2005sq}.
In the transverse and traceless gauge ($\nabla^a h_{ab} = g^{ab}h_{ab}=0$),
with $h_{ab}=0$ unless $(a,b)=(i,j)$ we can write
\be\label{generalperturbation}
h_{ij}(r,t,\mathbf{x})=r^2\,\phi(r,t)\,\bar{h}_{ij}(\mathbf{x})\,,
\ee%
where $\mathbf{x}$ is the direction of the sphere along which the perturbation propagates
and $\bar{h}_{ij}$ is the eigentensor of the Lichnerowicz operator built on the background
$3$-sphere
\be\label{LaplacianOperator}
\left(\bar{\bigtriangleup}_L+\g\right)\bar{h}_{ij}=0\,,\quad
\g=\ell(\ell+2)-2\,.
\ee%

The perturbations $h_{ij}$ are both gauge-invariant and
decouple~\cite{Dotti:2004sh,Dotti:2005sq,Ishibashi:2003ap,Kodama:2003jz,Gibbons:2002pq}.
This decoupling is a consequence of the spherical symmetry of the background
and occurs for every value of $\ell$ and not only in the hydrodynamic limit $\ell=\ell_0$.
Furthermore, assuming a harmonic time-dependence of the perturbation,
$h_i^j=\phi(r,t)\,\bar{h}_i^j(\mathbf{x})=\psi(r)\,e^{-i\omega t}\,\bar{h}_i^j(\mathbf{x})$,
the perturbation $h_i^j$ factorises leading to a set of
equations which depend only on $t$ and $r$~\cite{Dotti:2004sh,Dotti:2005sq}.
Thus \cref{perturbedGB} reduces to a massive scalar equation
\be\label{EquationsOfMotion}
\frac{1}{r^3}\frac{d}{dr}\left[r^3f(r)F(r)\frac{d\psi}{dr}\right]
+\omega^2\,\frac{F(r)}{f(r)}\,\psi - m^2(r)\,\psi=0\,,
\ee%
where $F(r)\equiv1-\l L^2 f'(r)/r$ and the mass term is
\be\label{massterm}
m^2(r) = \frac{2-\g}{r^2}\left[1-\l L^2 f''(r)\right]
+ T_{(M)}{}_i^i - \frac{\delta T_{(M)}{}_{ij}}{\delta g_{ij}}\,.
\ee%
 Notice that the mass term depends on
the angular part of the perturbation through the eigenvalue~$\g$
of the Lichnerowicz operator~\eqref{LaplacianOperator} and on higher-curvature
corrections through the GB constant~$\lambda$.
In the black brane case, if translational invariance is preserved,
the mass term is identically zero~\cite{Kovtun:2004de}.
We stress that, although \cref{EquationsOfMotion} holds for any $\ell$,
since we are interested in computing the shear viscosity~\eqref{analogueKubo},
in the following $\ell$ will be taken equal to its minimum value $\ell_0=1$
implying $\g=1$.

There are no general exact analytical solutions of \cref{EquationsOfMotion},
but approximate analytical solutions can be found for $r\to\infty$ and in the near-horizon limit.
In the generic case, solutions can only be computed numerically.

The asymptotic solutions of \cref{EquationsOfMotion} with $\omega=0$ are given in terms
of the modified Bessel functions of first and second kind. For $r\to\infty$,
the non-normalisable mode $\psi_0$ and the normalisable mode $\psi_1$ behave as
\be%
\psi_0 &= 1 - \frac{\l L^2}{2\left(1-\sqrt{1-4\l}\right) r^2}
+ \O\left(\log r/r^4\right),\label{asymptoticpsi0}\\
\psi_1 &= \frac{1}{r^4} + \O\left(1/r^6\right).
\ee%
In \cref{asymptoticpsi0} the integration constant is chosen such that the
non-normalisable mode $\psi_0$ goes to 1 as $r\to\infty$.

The near-horizon behaviour of $\psi_0(r)$ is different for non-extremal and extremal BHs.
In the case of non-extremal BHs at temperature $T$ and extremal $T=0$ BHs
we write the metric function, respectively
\be%
f(r) &= 4\pi T\,(r-r_+) + \frac{f''(r_+)}{2}\,(r-r_+)^2 + \O\left({(r-r_+)}^3\right),\label{cvq}\\
f(r) &= \frac{(r-r_0)^2}{l^2} + \O\left({(r-r_+)}^3\right),
\ee%
where the extremal BH radius $r_0$ is defined in \cref{horizonextreme}
and the $\AdS_2$ length $l$ is given by \cref{ads2length}.
In the non-extremal case, we write $\psi_0(r)$ using a power-series expansion,
and we solve \cref{EquationsOfMotion} order by order. At leading order we find:
\be%
\psi_0(r) = &~\psi_0(r_+)\left[ 1 + \frac{1-\l L^2 f''(r_+)}{4\pi T r_+^2 - \l L^2 r_+ (4\pi T)^2}\,(r-r_+)\right]\0\\
&+ \O\left({(r-r_+)}^2\right).\label{nearhorizonpsi}
\ee%
For the extremal case, the leading quadratic behaviour of $f(r)$ implies $\psi_0(r_+)=0$.
The behaviour of $\psi_0(r)$ in the near-horizon region is
\be\label{psi0nh-ext}
\psi_0(r) = (r-r_0)^\nu, \quad\nu= \frac{1}{2}\left(-1+\sqrt{1+ \frac{4 l^2-8\l L^2}{r_0^2}}\right).
\ee%

\section{The shear viscosity to entropy density ratio\label{sec:etatos}}

In this section, following the method proposed in Refs.~\cite{Hartnoll:2016tri,Lucas:2015vna},
we compute the shear viscosity to entropy ratio for the QFTs dual to
the BH solutions discussed in \cref{Sec:BHsol}. For Einstein gravity coupled to matter,
$\tilde\eta/s$ of the dual QFT is determined by means of the retarded Green
function in \cref{analogueKubo} and it is given by the non-normalisable mode~$\psi_0$
of the perturbation evaluated at the horizon,
\be%
\frac{\tilde\eta}{s}=\frac{1}{4\pi}\,\psi_0(r_+)^2\,.
\ee%
This method can be generalised to include higher-curvature contributions.
The computation uses a Wronskian method
to determine the relation between the normalisable mode $\psi_1$ and the non-normalisable
mode $\psi_0$. Since this relation does not depend on the mass term $m^2(r)$ in \cref{EquationsOfMotion},
the formula of Ref.~\cite{Hartnoll:2016tri} also holds for BHs in GB gravity:
\be%
\frac{\tilde\eta}{s} &= \frac{1}{4\pi}\,\psi_0(r_+)^2\,\times\0\\
&\left[1-4\l\left(1-\frac{2\pi G_5 Q^2 L^2}{3 r_+^6}\right)\right]
\left(1+\frac{6\l L^2}{r_+^2}\right)^{-1}\,,\label{etatos}
\ee%
where $\psi_0(r)$ is the non-normalisable solution of \cref{EquationsOfMotion} with $\omega=0$.

For background solutions which do not break translational invariance, \eg\ branes,
the mass term $m^2(r)$ is identically zero and the zero-frequency solution is
$\psi_0(r)=1$ everywhere~\cite{Hartnoll:2016tri,Cadoni:2016hhd}.
On the contrary, in BH backgrounds, the translational invariance is broken, the mass
term $m^2(r)$ is non-vanishing, the $\omega=0$ solution for $\psi_0(r)$ is not constant and $\psi_0(r_+)$
must be calculated by integrating \cref{EquationsOfMotion} with $\omega=0$.

Large radius BHs $r_+\gg L$, correspond to the large temperature regime $T\gg 1/L$.
In this approximation $T(r_+)$ can be inverted in \cref{MTS:temperature} to get
$r_+(T)= \pi L^2 T+ \O\left(1/T\right)$.
Then, using \cref{etatos,asymptoticpsi0} we get
\be\label{etatos:largeT}
\frac{\tilde\eta}{s}= \frac{1-4\l}{4\pi}
\left[1 - \frac{\l L^2 \left(7 -6\sqrt{1-4\l}\right)}{\pi^2\left(1-\sqrt{1-4\l}\right) L^4 T^2}
+ \O\left(1/T^4\right)\right].
\ee%
As expected, in the large~$T$ regime, $\tilde\eta/s$ does not depend on the charge.
For GR BHs, \cref{etatos:largeT} is a decreasing function of the temperature,
thus the KSS bound is violated and the universal value $1/4\pi$ is attained only for $T\to\infty$.
For GB BHs, the behaviour is qualitatively similar but as $T\to\infty$ the value of $\tilde\eta/s$
tends to $(1-4\l)/4\pi$.

In the extremal case, the metric function and its first derivative vanish when
evaluated on the horizon and following Ref.~\cite{Hartnoll:2016tri} the shear
viscosity to entropy ratio is given by
\be\label{etatosextreme}
\frac{\tilde\eta}{s}= \frac{1}{4\pi}\,\psi_0(r_+)^2 \left(1+\frac{6\l L^2}{r_0^2}\right)^{-1}.
\ee%
\Cref{psi0nh-ext} tells that $\psi_0(r_+)=0$, which substituted in \cref{etatosextreme}
means that $\tilde\eta/s$ goes to zero in the $T=0$ extremal limit.
The scaling at low temperatures of $\tilde\eta/s$ follows from simple matching
argument~\cite{Hartnoll:2016tri} between scaling of the Green function and the
near-horizon scaling~\eqref{psi0nh-ext}
\be\label{etatos-ext}
\frac{\tilde\eta}{s}\sim T^{2\nu}\,,
\ee%
where $\nu$ is given by~\eqref{psi0nh-ext}. The scaling exponent satisfies $\nu\le 1$ for
\be%
\l\ge\frac{l^4}{L^2r_0^2} +\frac{l^2}{2L^2}\,.
\ee%

The global behaviour of $\tilde\eta/s$ as
a function of $T$ is obtained by numerically integrating \cref{EquationsOfMotion}
supplied with a power-series boundary condition for $\psi_0(r)$.
In the following, we choose units $G_5=L=1$.
For each value of the charge and the GB parameter, there exists a minimum mass
(and hence a minimum radius) given by \cref{massbound}.
\Cref{EquationsOfMotion} is then integrated outwards from the horizon to infinity.
Next, a shooting method is used to determine $\psi_0(r_+)$ by requiring that $\psi_0(\infty)=1$.
Finally, the temperature and $\tilde\eta/s$ for each solution are computed with
\cref{MTS:temperature,etatos}.

\subsection{AdS-Reissner-Nordstr\"om black holes}

\begin{figure*}[!ht]
\centering%
\includegraphics[width=0.35\textwidth]{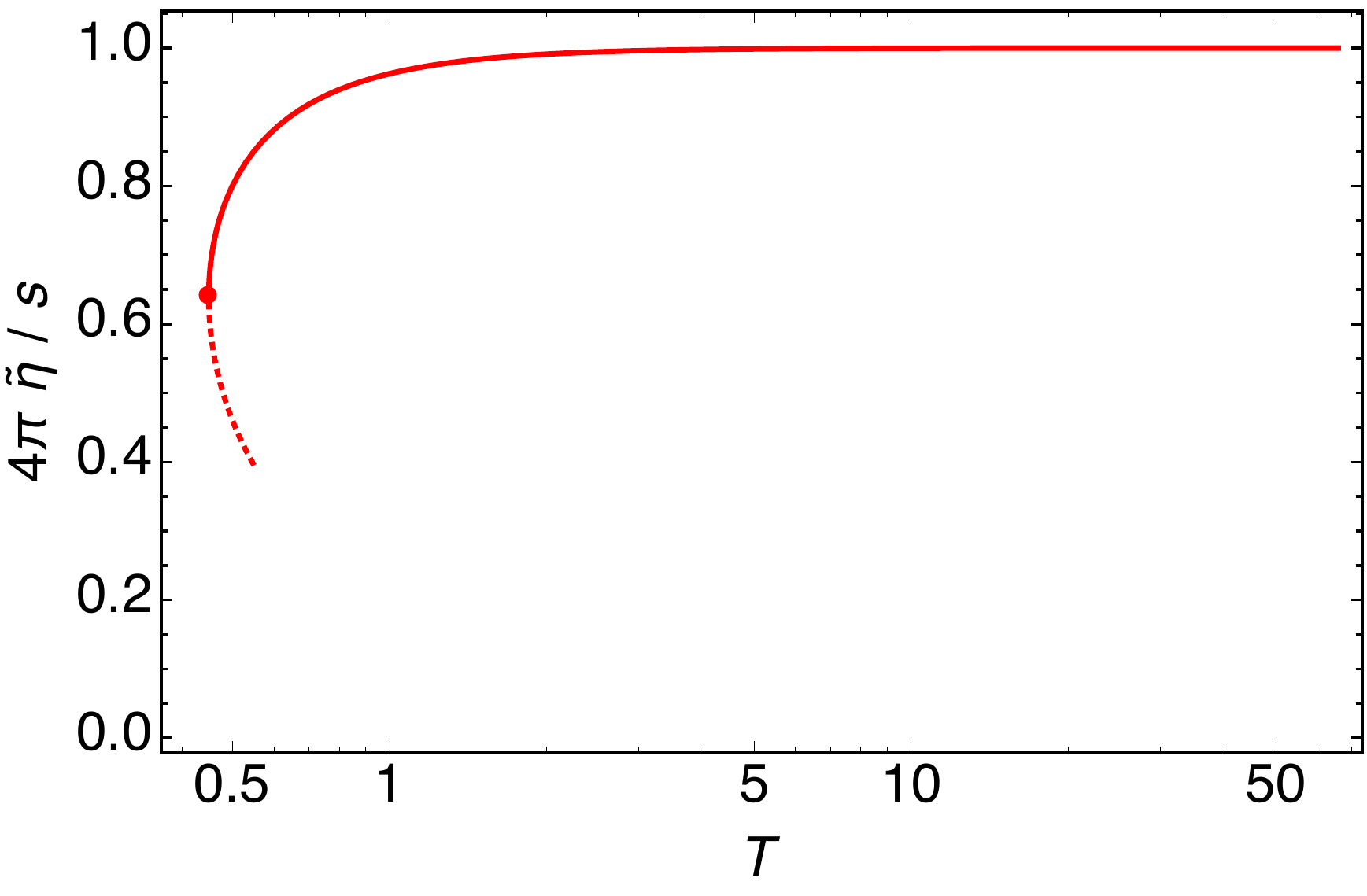}\qquad
\includegraphics[width=0.35\textwidth]{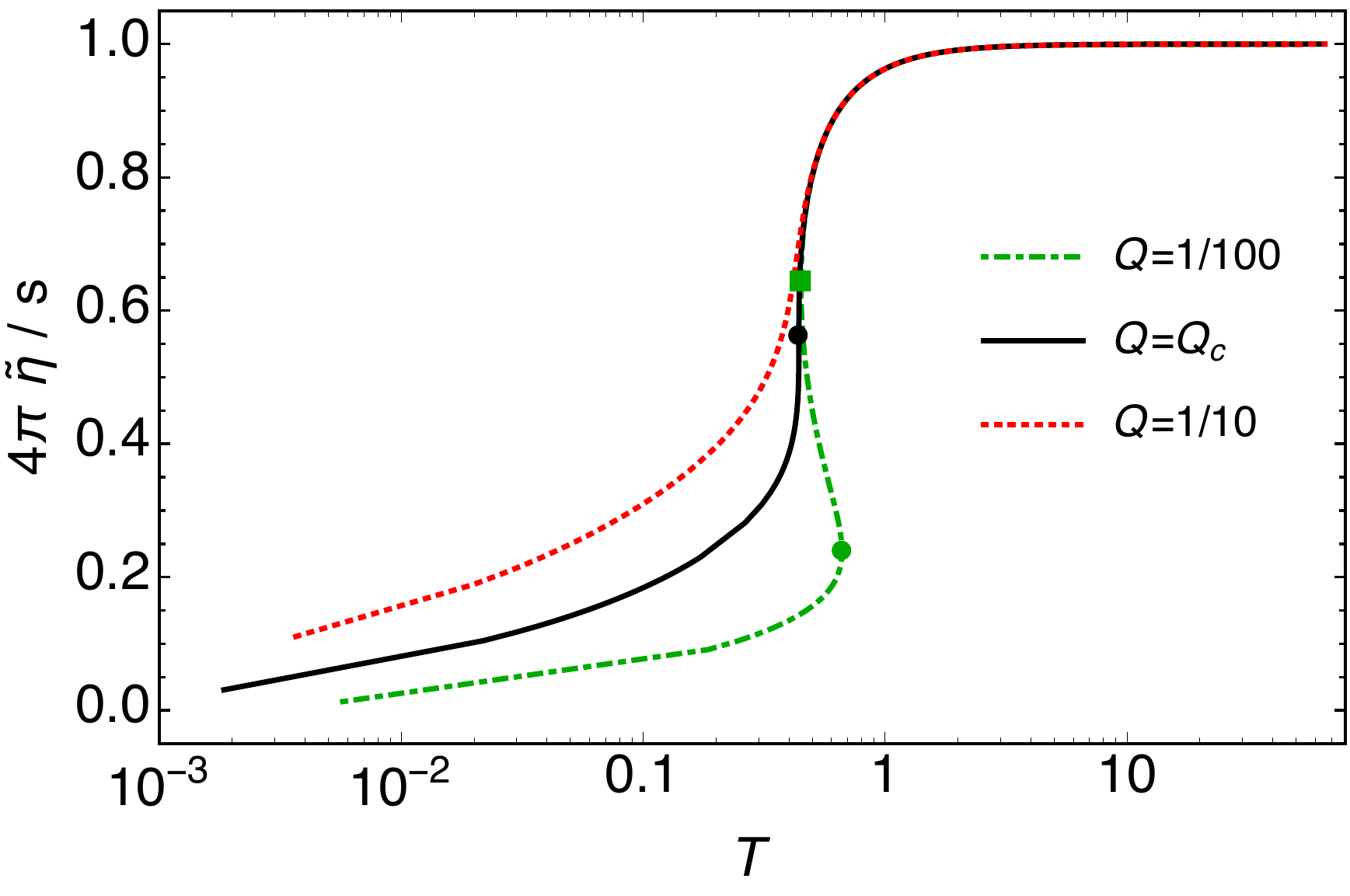}
\caption{Global behaviour of $\tilde{\eta}/s$ as a function of the temperature for GR BHs.
\emph{Left panel}: neutral AdS BHs. The solid line is the region above the critical radius,
while the dotted line represents (part of) the region below the critical radius,
where the BH is unstable and a thermal AdS solution is preferred; the dot marks the critical
radius at $T=\sqrt{2}/\pi$. \emph{Right panel}: AdS-RN BHs. We plot $\tilde{\eta}/s$ for three selected
values of the BH charge: above, at and below the critical value $Q_c=1/6\sqrt{5\pi}$, at which
the system undergoes the second-order phase transition. The dots (square) mark the maximum (minimum)
of the temperature as a function of the BH radius.}\label{fig:GR}
\end{figure*}

The plots of $\tilde\eta/s$ resulting from our numerical calculations for
GR are shown in \cref{fig:GR} for electrically
neutral (left panel) and charged (right panel) BHs.

The KSS bound is always violated for small and intermediate values of temperature,
whereas it is saturated from below for large temperatures.
In this section we extend the discussion of Ref.~\cite{Cadoni:2017fnd}.
For neutral AdS BHs, $\tilde{\eta}/s$ starts at the universal value $1/4\pi$ at
large temperatures and decreases monotonically as $T$ decreases, reaching a minimum
non-zero value for the non-vanishing minimum temperature $T=\sqrt{2}/\pi$.
Such a temperature corresponds to the minimum value of the BH radius, $r_0=1/\sqrt{2}$.
At $r=r_0$ there is the Hawking-Page transition and for $r_+\le r_0$ there are no stable
BH solutions~\cite{Birmingham:1998nr} and thermal AdS is energetically preferred with
respect to the BH\@. The dotted line in the left panel of \cref{fig:GR} gives
$\tilde{\eta}/s$ for BHs with radii less than $r_0$, whose behaviour is a consequence
of the growing of $T$ for \mbox{$r_+\le r_0$}.

For AdS-RN BHs, $\tilde{\eta}/s$ decreases from $1/4\pi$ at large temperatures
(independently from the charge), but the behaviour for small and intermediate
temperatures depends on the charge. As explained in \cref{subSec:AdSBHQ}, there
exists a critical value of the charge $Q_c=1/6\sqrt{5\pi}$ under which the system
undergoes a phase transition.
On the right panel of \cref{fig:GR}, we plot our numerical results for $\tilde{\eta}/s$
for the critical charge and for representative values of the charge above, at and
below the critical value.
The dots (squares) in the curves with $Q\leq Q_c$ mark the critical temperature $T_{\max}$
($T_{\min}$) corresponding, to the two local extrema of the function $T(r_+)$ of
\cref{MTS:temperature}. At these critical temperatures, the specific heat changes sign
according to the discussion in \cref{subSec:AdSBHQ}. For $Q=Q_c$ it is found that $T_{\min}=T_{\max}$ and the
function $T(r_+)$ has an inflection point. For $Q>Q_c$ the function $T(r_+)$ is monotonically increasing
and BHs are always stable. The numerical values $T_{\min}$ and $T_{\max}$ are listed in \cref{tab}
for a representative value of the charge below and at $Q_c$.

Interestingly, $\tilde\eta/s$ develops hysteresis for $0<Q<Q_c$
This is evident for the $Q=1/100$ solid black curve in the right panel of \cref{fig:GR}.
We have also checked that curves with $Q<Q_c$ have a similar hysteretic behaviour,
whereas those with $Q>Q_c$ (as the $Q=1/10$ orange dashed line) do not show this feature.
Notice that the limit $Q\to 0$ in the plot of l.h.s.\ of \cref{fig:GR} is singular. 
As explained in \cref{subSec:AdSBHQ}, we have a discontinuity for $Q\to0$, \ie\ 
there is no phase transition at $Q=0$. Because the existence of the phase transition 
is a necessary condition for having hysteresis in $\tilde\eta/s$, this means that also 
$\tilde\eta/s$ as a function of the temperature is discontinuous at $Q=0$: 
there is a more pronounced hysteretical behaviour for $Q\to 0$, but 
hysteresis disappears completely at $Q=0$.
This hysteretic behaviour is a direct consequence of
the Van der Waals-like behaviour of the AdS-RN BHs discussed in \cref{subSec:AdSBHQ}. It is
related to the presence of two local extrema in the function $T(r_+)$ in \cref{MTS:temperature}
or equivalently, to the presence of two stable states (small and large BHs) connected by
a meta-stable region (intermediate BHs). 
This phase portrait has been considered as a general explanation of hysteretic
behaviour for some variables of the system~\cite{Knittel:1977}.
In particular, when the system evolves from high (low) to lower (higher) temperatures,
a potential barrier prevents the evolution of the system from occurring as an
equilibrium path between the two stable states~\cite{Bertotti:1998}.
Equilibrium will be reached passing through a meta-stable region and a path-dependence
of $\tilde\eta/s$ is generated. In particular, starting from high temperatures, the
system will reach low temperatures going directly from the minimum and vice-versa.
The presence of these local extrema determines the patterns of signs of the BH specific
heat and free energy, hence the local thermodynamical stability~\cite{Chamblin:1999tk,Cai:2001dz}.
Thus, hysteresis in $\tilde\eta/s$ and thermodynamical phase transition have the same
origin and pattern. In fact, as already noted in \cref{subSec:AdSBHQ}, the phase diagram
of AdS-RN BHs is very similar to that of a Van der Waals liquid/gas transition.

This is a very interesting result: $\tilde\eta/s$ for the dual QFT carries direct
information about the thermodynamic phase transitions of the system.
In the holographic context, a hysteretic behaviour in the shear viscosity has been
already observed in Ref.~\cite{Erdmenger:2010xm, Erdmenger:2011tj} for AdS BHs with broken rotational
symmetry and with a p-wave holographic superfluid dual. Moreover, it is well-known that
nanofluids may exhibit hysteresis in the $\eta$-$T$ plane~\cite{Nguyen:2008}.

Notice that, even though solutions with $Q>Q_c$ describe stable BHs in
the overall range of $T$, our numerical computation does not hold in the small $T$ regime
as it uses a power-series near-horizon expansion.
However, $\tilde\eta/s\to0$ as $T\to0$ with analytical scaling law~\eqref{etatos-ext}
and scaling exponent $\nu$ given by \cref{psi0nh-ext} with $\l=0$.

\subsection{Neutral Gauss-Bonnet black holes}

\begin{figure}[!ht]
\centering
\includegraphics[width=0.4\textwidth]{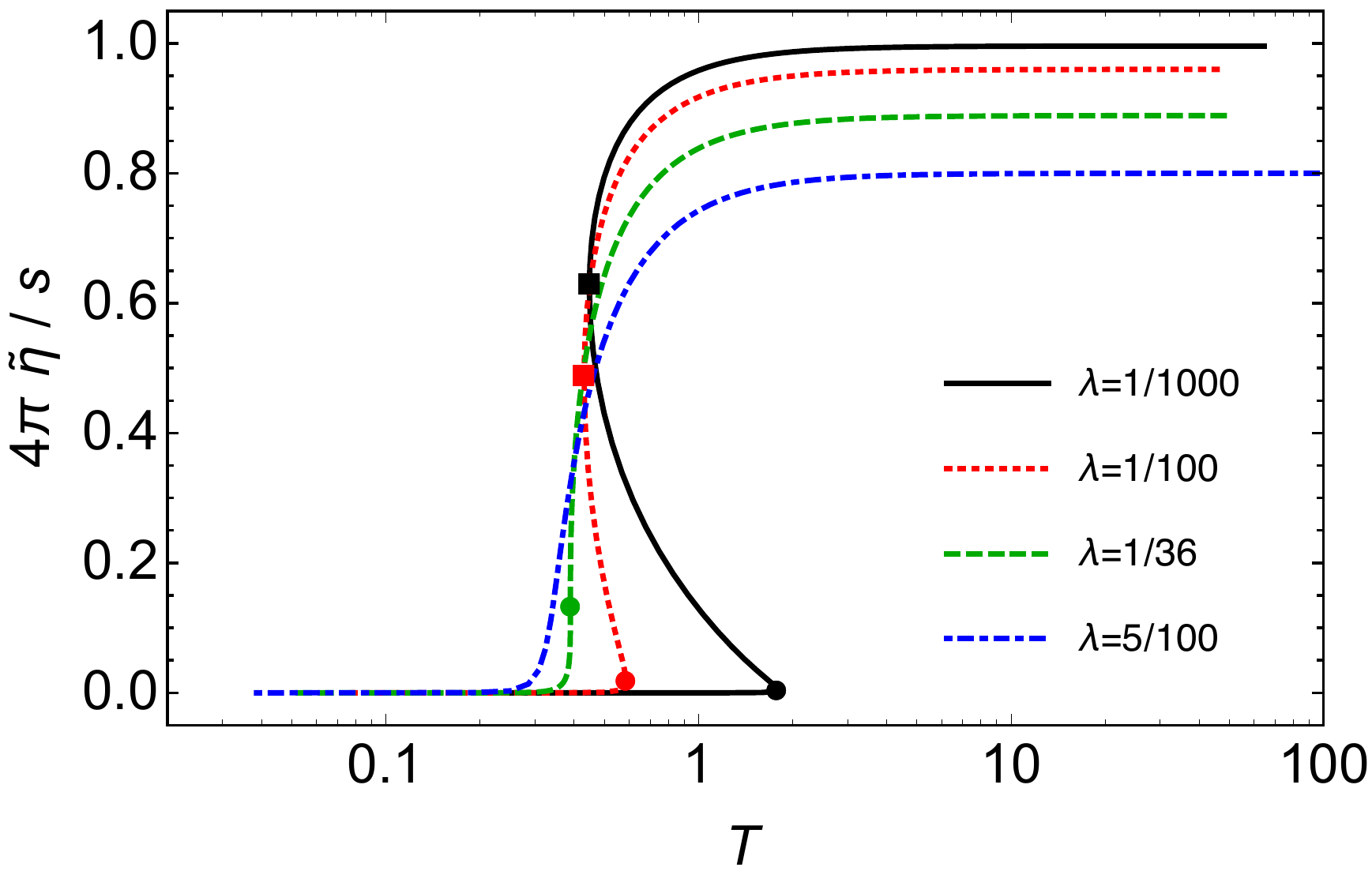}
\caption{Global behaviour of $\tilde\eta/s$ as a function of the temperature for
GB BHs with $Q=0$ for selected values of the GB coupling constant above, at and
below the critical value. Dots (squares) mark the maximum (minimum) of the
temperature as a function of the BH radius.}
\label{fig:GBQ0}
\end{figure}

Our numerical results for $\tilde\eta/s$ as a function of $T$ for neutral GB BHs
are shown in \cref{fig:GBQ0} for selected values of the
GB parameter $\l$ in the range $0<\l\le5/100$.

For large temperatures, the KSS bound is always violated due to the GB
contribution and $4\pi\tilde\eta/s\to1-4\l$.
At intermediate temperatures, the behaviour is qualitatively similar to that of RN BHs,
with the GB parameter $\l$ playing the role of the charge $Q$. As discussed in \cref{section:ngbbh},
there exists a critical value $\l_c$ under which GB BHs can undergo a phase transition:
by numerical investigation this value is $\l_c=1/36$, in good agreement with
Refs.~\cite{Zeng:2016aly,Cai:2001dz}.
Curves with $0<\l<\l_c$ (black solid and red dotted lines) show a hysteretic behaviour
of $\tilde\eta/s$ as a function of the temperature, whereas those with $\l>\l_c$ do not.
For a given value $\l<\l_c$, there are two critical temperatures $T_{\max},T_{\min}$,
which are marked respectively by dots and squares in the curves of \cref{fig:GBQ0}.
Their numerical values for selected values of $\l$ are listed in \cref{tab}.
Notice that similarly to the $Q\to0$ case, the limit $\l=0$ in the plots of \cref{fig:GBQ0} is singular. 
As explained in \cref{section:ngbbh}, for $\l=0$ there is a discontinuity. This implies 
that also $\tilde\eta/s$ as a function of the temperature is discontinuous at $\lambda=0$.  
There is a more pronounced hysteretical behaviour for smaller and smaller values of $\lambda$,
but hysteresis disappears completely at $\lambda=0$.

The physical interpretation of the appearance of hysteresis in $\tilde\eta/s$ for
the QFT dual to the neutral GB BH is completely analogue to that discussed for the
AdS-RN BH\@. When $\l$ reaches the critical value, the system undergoes a second-order
Van der Waals-like phase transition and exhibits the hysteretic behaviour in $\tilde\eta/s$.

\subsection{Charged Gauss-Bonnet black holes}

The presence of both a non-vanishing charge and GB coupling constant makes
the case of charged GB BHs more involved. However, as discussed in \cref{section:chargedbh},
the phase portrait becomes much simpler and has a Van der Waals-like form if we restrict
our considerations to the region where BHs are globally stable and either holds $Q$ or $\l$ fixed.
In this situation we expect the qualitative behaviour of $\tilde\eta/s$ as a function of $T$ to be
quite similar to that found for the AdS-RN and the neutral GB BHs.
\begin{figure*}[!ht]
\centering
\includegraphics[height=0.25\textwidth]{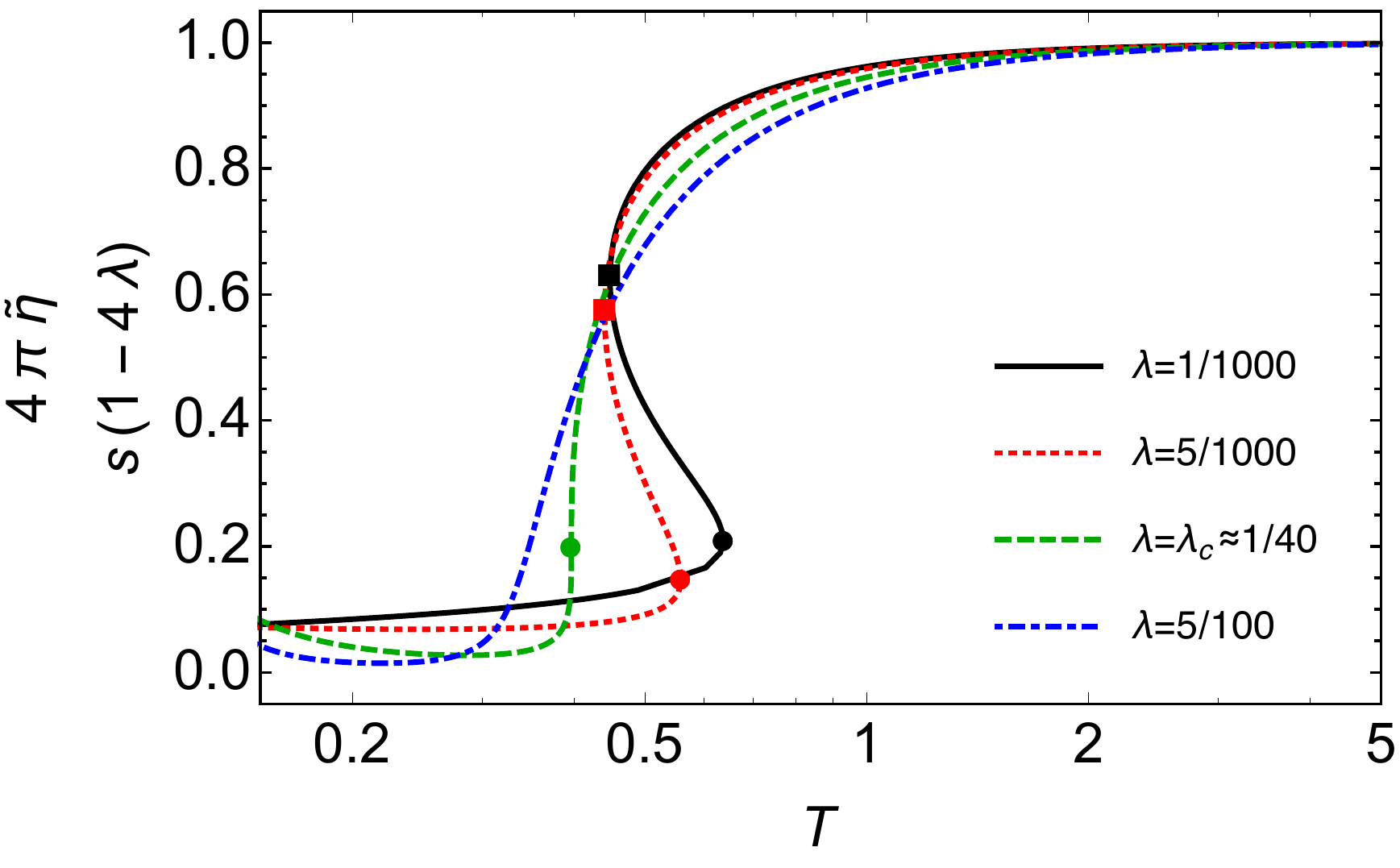}\qquad
\includegraphics[height=0.25\textwidth]{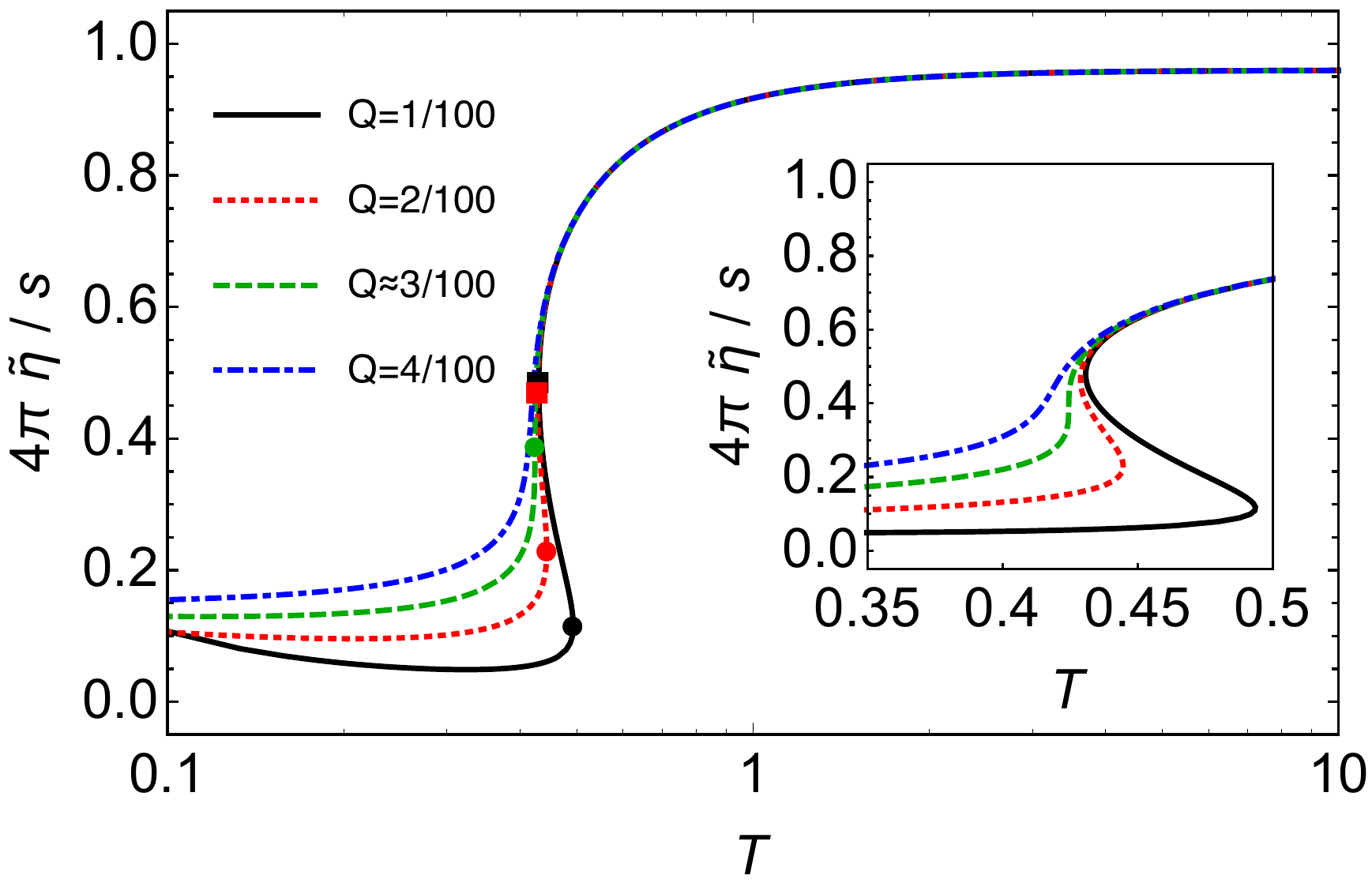}
\caption{Global behaviour of $\tilde\eta/s$ as a function of the temperature for
charged GB BHs. \emph{Left panel}: GB BHs with fixed charge $Q=1/100$, and selected
values of GB constant $\l=1/1000,\ 5/1000,\ \sim0.25,\ 5/100$. The value of
$\tilde\eta/s$ is rescaled by a factor $1-4\l$; in this way the large~$T$
behaviour of $4\pi\tilde\eta/s$, which for GB gravity is $\l$-dependent, has been
normalised to $1$. \emph{Right panel}: GB BHs with fixed value of GB constant
$\l=1/100$, and selected values of charge $Q=1/100,\ 2/100,\ \sim3/100,\ 4/100$.
Inset: zoom of the hysteresis region. Dots (squares) mark the local maximum
$T_{\max}$ (local minimum $T_{\min}$) of the temperature.}\label{fig:GB}
\end{figure*}
The numerical results for $\tilde\eta/s$ as a function of $T$, confirm our expectation
and are shown in \cref{fig:GB} for $Q$ fixed and selected values of the GB parameter
$\l$ (left panel) and for $\l$ fixed and selected values of the charge $Q$ (right
panel). In both cases, the numerical results corroborate the analytical ones.
For large temperatures, the KSS bound is always violated as $4\pi\tilde\eta/s\to1-4\l$.
At intermediate temperatures, the behaviour of $\tilde\eta/s$ depends crucially
on the values of the parameters $Q$ and $\l$.
For large values of $Q$ (or for values of $\l$ near to the unitarity bound
$\l\lesssim9/100$), large BHs are always stable, $\tilde\eta/s$ decreases monotonically
with $T$ and there is no hysteresis.

Notice that the limits $\l\to 0$, respectively $Q\to 0$, are singular in the plot 
on the left, respectively on the right, of Fig. \ref{fig:GB}.  Here we have a 
discontinuous behaviour of $\tilde\eta/s$ similar to that found in the 
$Q\to 0$ limit for charged BHs in GR and to the 
$\l\to 0$ for the uncharged BHs of GB gravity.

The situation changes drastically for $Q$ (or $\l$) of order $3/100$ and smaller:
the system may undergo a Van der Waals-like phase transition. The function $T(r_+)$
develops two local extrema $T_{\min}$ and $T_{\max}$, signalising the presence of two
different stable thermodynamical phase (small and large BHs) connected by a meta-stable one,
correspondingly, the $\tilde\eta/s$ curve as a function of $T$ develops hysteresis.
Two typical examples of this hysteretic behaviour are shown in \cref{fig:GB}.
On the left panel, for fixed $Q=1/100$, the onset of hysteresis can be seen, corresponding to
the thermodynamical phase transition, when $\l\lesssim0.025$.
On the right panel, for fixed $\l=1/100$, we see the onset of
hysteresis and the thermodynamical phase transition when $Q\lesssim3/100$.
The corresponding values of the critical temperatures are marked by the dots ($T_{\max}$)
and squares ($T_{\min}$) in \cref{fig:GB} and their numerical values are listed in \cref{tab}
for selected values of the parameters $Q$ and $\l$.
Analogous results can be found by choosing different $Q$ and $\l$.
Notice that, for stable BH solutions with values of $\l$ and $Q$ above the critical
values, our numerical computation cannot reach $T\sim 0$, because it uses a power-series
near-horizon expansion which does not hold in the extremal case. However,
from \cref{etatos-ext,psi0nh-ext}, which describe analytically
the near-extremal behaviour we conclude that $\tilde\eta/s\to0$ smoothly as $T\to0$.

\begin{table*}[!ht]
\centering%
\begin{tabular}{cccccccccccc}
\toprule
$Q$ & 1/100 & $Q_c$ & \multicolumn{3}{c}{0} & \multicolumn{3}{c}{1/100} & 1/100 & 2/100 & $Q_c$ \\\midrule
$\l$ & \multicolumn{2}{c}{0} & 1/1000 & 1/100 & $\l_c$ & 1/1000 & 5/1000 & $\l_c$ & \multicolumn{3}{c}{1/100}\\\midrule
$T_{\min}$ & 0.449 & 0.441 & 0.448 & 0.431 & 0.390 & 0.448 & 0.440 & 0.397 & 0.431 & 0.429 & 0.425 \\
$T_{\max}$ & 0.664 & 0.441 & 1.787 & 0.587 & 0.390 & 0.638 & 0.559 & 0.397 & 0.494 & 0.445 & 0.425 \\\bottomrule
\end{tabular}
\caption{Critical temperatures for selected values of $\l$ and $Q$ below and at the critical values.
For the AdS-RN BH the critical charge is $Q_c=1/6\sqrt{5\pi}$.
For the neutral GB BH the critical value of the coupling is $\l_c=1/36$.
For GB BHs with fixed charge $Q=1/100$ the critical value of the coupling is $\l_c\approx1/4$,
while for fixed $\l=1/100$ the critical charge is $Q_c\approx3/100$.}\label{tab}
\end{table*}

\section{Summary and outlook\label{sec:conclusions}}

In this paper we have used the AdS/CFT correspondence to obtain information about
the behaviour of bulk BHs by studying the hydrodynamic properties of the dual QFTs.
In particular, we have defined and computed the shear viscosity to entropy ratio in the transverse channel
for QFTs holographically dual to five-dimensional AdS BH solutions of GR and GB gravity.
In this way, we have extended the usual derivation of
$\tilde\eta/s$ for QFTs dual to gravitational bulk backgrounds with planar horizons
to backgrounds with spherical horizons.
We have shown that in holographic models the shear viscosity to entropy
ratio of the QFT is closely related and keeps detailed information about the
thermodynamical phase structure of the dual BH background.
This is not completely unexpected, because experience with another holographic
condensed matter system, like holographic superconductors, has shown us that
transport features of the dual QFT may be strongly related to phase transitions
of the dual black brane.

In general, the definition of a transport coefficient such as the shear viscosity is
associated with the translational invariance of the system, \ie\ the conservation
of the momentum. As a consequence the Fick law of diffusion can be derived
from the associated conserved current.
For systems that break translational invariance, the hydrodynamic interpretation
in terms of conserved quantities fails but hydrodynamics can be still
defined as an expansion in the derivatives of the hydrodynamic fields.
In this way, it is possible to define the shear viscosity through a Kubo
formula also for QFTs on a spherical background, see \cref{analogueKubo},
where the stress-energy tensor is only covariantly conserved.
In addition, one can understand $\tilde\eta$ as the rate of entropy production
due to a strain, which is the typical interpretation when
the homogeneity is broken by external matter fields.
From this point of view, QFTs dual to spherical BHs are very similar to QFTs dual
to black branes where the translational symmetry is broken by non-homogeneous
external fields, \eg\ scalars~\cite{Hartnoll:2016tri,Burikham:2016roo,Alberte:2016xja}.

The definition of the hydrodynamic limit of a QFT on the sphere is plagued by an
issue related to the compactness of the space. In fact, in a compact space, the
usual hydrodynamic limit as an effective theory describing the long-wavelength modes
of the QFT has not a straightforward interpretation.
Our proposal is that for QFTs dual to bulk spherical BHs, the hydrodynamical,
long wavelength modes can be described by the $\ell\to\ell_0$ modes that probe
large angles on the sphere.
This is in analogy with the $k\to0$ modes for QFTs dual to bulk black branes
which probe large scales on the plane.

There is still a crucial difference between the two cases.
When the breaking of translational symmetry is generated by external fields,
the symmetry may be restored or not when the system flows to the IR~\cite{Hartnoll:2016tri}.
Instead, in the BH case, because the breaking has a geometric and topological
origin, translational symmetry cannot be restored in the IR\@.

As expected, the large~$T$ behaviour of $\tilde\eta/s$, corresponding to the flow to the UV fixed point,
reproduces the universal value $1/4\pi$ or $(1-4\l)/4\pi$ in the GB case.
When the bulk BH solution has a regular and stable extremal limit (like \eg\ charged BHs)
and remains stable at small $T$, $\tilde\eta/s\to0$ as $T\to0$ with a $T^{2\nu}$ scaling law.
In the latter case, the system flows in the IR to the $\AdS_2\times S^3$ geometry.

Our most important result is the behaviour of $\tilde\eta/s$ at intermediate temperatures.
A second-order, Van der Waals-like, phase transition 
occurs when the control parameters go below their critical values~\cite{Chamblin:1999tk,Chamblin:1999hg}. 
In this situation BHs may also undergo a first order phase transition 
controlled by the temperature. This corresponds to the transition 
from small to large BHs connected through a meta-stable intermediate 
region. As a consequence, $\tilde\eta/s$ as a function of $T$ 
always develops hysteresis and it becomes multi-valued as expected 
for a first order phase transition~\cite{Erdmenger:2010xm}. Notice 
that in this case the first and second-order phase transitions are 
both necessary in order to have the hysteretic behaviour in $\eta/s$. 
Even though, similarly to the case discussed in Ref.~\cite{Erdmenger:2010xm} 
the multi-valuedness of $\eta/s$ is directly related only to the 
first order one. The role of the second-order phase transition is 
to allow for the existence of the first order one.

The mechanism that generates hysteresis in $\tilde\eta/s$ is the same that
is responsible for the phase transition and can be traced back to non-equilibrium
thermodynamics. When a control parameter, \ie\ the charge $Q$ or the GB coupling
constant $\l$, is below its critical value, the function $T(r_+)$ develops both
a local maximum and minimum. The regions below the maximum and above the minimum
correspond to two stable solutions, \ie\ small and large BHs, respectively.
The region between these two is represented by an unstable (meta-stable) region
of intermediate BHs. When the system evolves from large (small) BHs to small (large)
BHs, a potential barrier prevents the evolution of the system from occurring as
an equilibrium path between the two stable states~\cite{Bertotti:1998}.
Equilibrium will be reached passing through a meta-stable region~\cite{Knittel:1977},
and a path-dependence of $\tilde\eta/s$ is generated. The presence of these local
extrema determines the patterns of signs of the BH specific heat and free energy,
hence the local thermodynamical stability~\cite{Chamblin:1999tk,Cai:2001dz}.
This interesting result represents the first attempt to infer about BH
thermodynamics through a detailed analysis of a transport coefficient as the shear
viscosity.

Our definition of $\tilde \eta$ for spherical backgrounds is channel-dependent. In
general we have three different determinations of $\tilde \eta$ for shear, sound and
transverse (scalar) perturbations. In this paper we have focused on transverse perturbations.
It would be of interest to check whether the behaviour of the viscosity found in this paper
for the transverse channel also extends to the sound and shear channels. The computation of
our analogue $\tilde \eta$ in these other two channels is rather involved and we have left
it for future investigations.

\bibliographystyle{spphys.bst}
\pdfbookmark{\refname}{References}

\begin{thebibliography}{10}
\providecommand{\url}[1]{{#1}}
\providecommand{\urlprefix}{URL }
\expandafter\ifx\csname urlstyle\endcsname\relax
  \providecommand{\doi}[1]{DOI \discretionary{}{}{}#1}\else
  \providecommand{\doi}{DOI \discretionary{}{}{}\begingroup
  \urlstyle{rm}\Url}\fi

\bibitem{Policastro:2001yc}
G.~Policastro, D.T. Son, A.O. Starinets, Phys. Rev. Lett. \textbf{87}, 081601
  (2001).
\newblock \doi{10.1103/PhysRevLett.87.081601}

\bibitem{Buchel:2004qq}
A.~Buchel, Phys. Lett. \textbf{B609}, 392 (2005).
\newblock \doi{10.1016/j.physletb.2005.01.052}

\bibitem{Benincasa:2006fu}
P.~Benincasa, A.~Buchel, R.~Naryshkin, Phys. Lett. \textbf{B645}, 309 (2007).
\newblock \doi{10.1016/j.physletb.2006.12.030}

\bibitem{Kats:2007mq}
Y.~Kats, P.~Petrov, J. High Energy Phys. \textbf{01}, 044 (2009).
\newblock \doi{10.1088/1126-6708/2009/01/044}

\bibitem{Landsteiner:2007bd}
K.~Landsteiner, J.~Mas, J. High Energy Phys. \textbf{07}, 088 (2007).
\newblock \doi{10.1088/1126-6708/2007/07/088}

\bibitem{Iqbal:2008by}
N.~Iqbal, H.~Liu, Phys. Rev. \textbf{D79}, 025023 (2009).
\newblock \doi{10.1103/PhysRevD.79.025023}

\bibitem{Buchbinder:2008nf}
E.I. Buchbinder, A.~Buchel, Phys. Rev. \textbf{D79}, 046006 (2009).
\newblock \doi{10.1103/PhysRevD.79.046006}

\bibitem{Edalati:2009bi}
M.~Edalati, J.I. Jottar, R.G. Leigh, J. High Energy Phys. \textbf{01}, 018
  (2010).
\newblock \doi{10.1007/JHEP01(2010)018}

\bibitem{Kovtun:2003wp}
P.~Kovtun, D.T. Son, A.O. Starinets, J. High Energy Phys. \textbf{10}, 064
  (2003).
\newblock \doi{10.1088/1126-6708/2003/10/064}

\bibitem{Kovtun:2004de}
P.~Kovtun, D.T. Son, A.O. Starinets, Phys. Rev. Lett. \textbf{94}, 111601
  (2005).
\newblock \doi{10.1103/PhysRevLett.94.111601}

\bibitem{Song:2010mg}
H.~Song, S.A. Bass, U.~Heinz, T.~Hirano, C.~Shen, Phys. Rev. Lett.
  \textbf{106}, 192301 (2011).
\newblock \doi{10.1103/PhysRevLett.106.192301}.
\newblock (Erratum: ibid.\ {\bf 109}, 139904 (2012))

\bibitem{Strominger:1996sh}
A.~Strominger, C.~Vafa, Phys. Lett. \textbf{B379}, 99 (1996).
\newblock \doi{10.1016/0370-2693(96)00345-0}

\bibitem{Cadoni:1998sg}
M.~Cadoni, S.~Mignemi, Phys. Rev. \textbf{D59}, 081501 (1999).
\newblock \doi{10.1103/PhysRevD.59.081501}

\bibitem{Chamblin:1999tk}
A.~Chamblin, R.~Emparan, C.V. Johnson, R.C. Myers, Phys. Rev. \textbf{D60},
  064018 (1999).
\newblock \doi{10.1103/PhysRevD.60.064018}

\bibitem{Chamblin:1999hg}
A.~Chamblin, R.~Emparan, C.V. Johnson, R.C. Myers, Phys. Rev. \textbf{D60},
  104026 (1999).
\newblock \doi{10.1103/PhysRevD.60.104026}

\bibitem{Cvetic:2001bk}
M.~Cveti\v{c}, S.~Nojiri, S.D. Odintsov, Nucl. Phys. \textbf{B628}, 295 (2002).
\newblock \doi{10.1016/S0550-3213(02)00075-5}

\bibitem{Brigante:2007nu}
M.~Brigante, H.~Liu, R.C. Myers, S.~Shenker, S.~Yaida, Phys. Rev. \textbf{D77},
  126006 (2008).
\newblock \doi{10.1103/PhysRevD.77.126006}

\bibitem{Brigante:2008gz}
M.~Brigante, H.~Liu, R.C. Myers, S.~Shenker, S.~Yaida, Phys. Rev. Lett.
  \textbf{100}, 191601 (2008).
\newblock \doi{10.1103/PhysRevLett.100.191601}

\bibitem{Ge:2009ac}
X.H. Ge, S.J. Sin, S.F. Wu, G.H. Yang, Phys. Rev. \textbf{D80}, 104019 (2009).
\newblock \doi{10.1103/PhysRevD.80.104019}

\bibitem{Ge:2009eh}
X.H. Ge, S.J. Sin, J. High Energy Phys. \textbf{05}, 051 (2009).
\newblock \doi{10.1088/1126-6708/2009/05/051}

\bibitem{Cai:2009zv}
R.G. Cai, Z.Y. Nie, N.~Ohta, Y.W. Sun, Phys. Rev. \textbf{D79}, 066004 (2009).
\newblock \doi{10.1103/PhysRevD.79.066004}

\bibitem{Camanho:2010ru}
X.O. Camanho, J.D. Edelstein, M.F. Paulos, J. High Energy Phys. \textbf{05},
  127 (2011).
\newblock \doi{10.1007/JHEP05(2011)127}

\bibitem{Cremonini:2011iq}
S.~Cremonini, Mod. Phys. Lett. \textbf{B25}, 1867 (2011).
\newblock \doi{10.1142/S0217984911027315}

\bibitem{Jacobson:2011dz}
T.~Jacobson, A.~Mohd, S.~Sarkar, Phys. Rev. \textbf{D95}, 064036 (2011).
\newblock \doi{10.1103/PhysRevD.95.064036}

\bibitem{Bhattacharyya:2014wfa}
A.~Bhattacharyya, D.~Roychowdhury, J. High Energy Phys. \textbf{03}, 063
  (2015).
\newblock \doi{10.1007/JHEP03(2015)063}

\bibitem{Sadeghi:2015vaa}
M.~Sadeghi, S.~Parvizi, Class. Quantum Grav. \textbf{33}, 035005 (2016).
\newblock \doi{10.1088/0264-9381/33/3/035005}

\bibitem{Wang:2016vmm}
Y.L. Wang, X.H. Ge, Phys. Rev. \textbf{D94}(6), 066007 (2016).
\newblock \doi{10.1103/PhysRevD.94.066007}

\bibitem{Cadoni:2016hhd}
M.~Cadoni, A.M. Frassino, M.~Tuveri, J. High Energy Phys. \textbf{05}, 101
  (2016).
\newblock \doi{10.1007/JHEP05(2016)101}

\bibitem{Erdmenger:2010xm}
J.~Erdmenger, P.~Kerner, H.~Zeller, Phys. Lett. \textbf{B699}, 301 (2011).
\newblock \doi{10.1016/j.physletb.2011.04.009}

\bibitem{Erdmenger:2011tj}
J.~Erdmenger, P.~Kerner, H.~Zeller, J. High Energy Phys. \textbf{01}, 059
  (2012).
\newblock \doi{10.1007/JHEP01(2012)059}

\bibitem{Rebhan:2011vd}
A.~Rebhan, D.~Steineder, Phys. Rev. Lett. \textbf{108}, 021601 (2012).
\newblock \doi{10.1103/PhysRevLett.108.021601}

\bibitem{Mamo:2012sy}
K.A. Mamo, J. High Energy Phys. \textbf{10}, 070 (2012).
\newblock \doi{10.1007/JHEP10(2012)070}

\bibitem{Davison:2015taa}
R.A. Davison, B.~Gout\'eraux, S.A. Hartnoll, J. High Energy Phys. \textbf{10},
  112 (2015).
\newblock \doi{10.1007/JHEP10(2015)112}

\bibitem{Hartnoll:2016tri}
S.A. Hartnoll, D.M. Ramirez, J.E. Santos, J. High Energy Phys. \textbf{03}, 170
  (2016).
\newblock \doi{10.1007/JHEP03(2016)170}

\bibitem{Burikham:2016roo}
P.~Burikham, N.~Poovuttikul, Phys. Rev. \textbf{D94}, 106001 (2016).
\newblock \doi{10.1103/PhysRevD.94.106001}

\bibitem{Alberte:2016xja}
L.~Alberte, M.~Baggioli, O.~Pujolas, J. High Energy Phys. \textbf{07}, 074
  (2016).
\newblock \doi{10.1007/JHEP07(2016)074}

\bibitem{Liu:2016njg}
H.S. Liu, H.~Lu, C.N. Pope, J. High Energy Phys. \textbf{12}, 097 (2016).
\newblock \doi{10.1007/JHEP12(2016)097}

\bibitem{Buchel:2004di}
A.~Buchel, J.T. Liu, A.O. Starinets, Nucl. Phys. \textbf{B707}, 56 (2005).
\newblock \doi{10.1016/j.nuclphysb.2004.11.055}

\bibitem{Buchel:2008vz}
A.~Buchel, R.C. Myers, A.~Sinha, J. High Energy Phys. \textbf{03}, 084 (2009).
\newblock \doi{10.1088/1126-6708/2009/03/084}

\bibitem{Buchel:2009tt}
A.~Buchel, R.C. Myers, J. High Energy Phys. \textbf{08}, 016 (2009).
\newblock \doi{10.1088/1126-6708/2009/08/016}

\bibitem{Hofman:2009ug}
D.M. Hofman, Nucl. Phys. \textbf{B823}, 174 (2009).
\newblock \doi{10.1016/j.nuclphysb.2009.08.001}

\bibitem{Romatschke:2009kr}
P.~Romatschke, Class. Quantum Grav. \textbf{27}, 025006 (2010).
\newblock \doi{10.1088/0264-9381/27/2/025006}

\bibitem{Cremonini:2012ny}
S.~Cremonini, U.~G{\"u}rsoy, P.~Szepietowski, J. High Energy Phys. \textbf{08},
  167 (2012).
\newblock \doi{10.1007/JHEP08(2012)167}

\bibitem{Cremonini:2011ej}
S.~Cremonini, P.~Szepietowski, J. High Energy Phys. \textbf{02}, 038 (2012).
\newblock \doi{10.1007/JHEP02(2012)038}

\bibitem{Cadoni:2017fnd}
M.~Cadoni, E.~Franzin, M.~Tuveri, Phys. Lett. \textbf{B768}, 393 (2017).
\newblock \doi{10.1016/j.physletb.2017.02.060}

\bibitem{Ciobanu:2017fef}
T.~Ciobanu, D.M. Ramirez, arXiv:1708.04997  (2017)

\bibitem{Lucas:2015vna}
A.~Lucas, J. High Energy Phys. \textbf{03}, 071 (2015).
\newblock \doi{10.1007/JHEP03(2015)071}

\bibitem{Baier:2007ix}
R.~Baier, P.~Romatschke, D.T. Son, A.O. Starinets, M.A. Stephanov, J. High
  Energy Phys. \textbf{04}, 100 (2008).
\newblock \doi{10.1088/1126-6708/2008/04/100}

\bibitem{Kovtun:2012rj}
P.~Kovtun, J. Phys. \textbf{A45}, 473001 (2012).
\newblock \doi{10.1088/1751-8113/45/47/473001}

\bibitem{Son:2007vk}
D.T. Son, A.O. Starinets, Ann. Rev. Nucl. Part. Sci. \textbf{57}, 95 (2007).
\newblock \doi{10.1146/annurev.nucl.57.090506.123120}

\bibitem{Gubser:1998bc}
S.S. Gubser, I.R. Klebanov, A.M. Polyakov, Phys. Lett. \textbf{B428}, 105
  (1998).
\newblock \doi{10.1016/S0370-2693(98)00377-3}

\bibitem{Witten:1998qj}
E.~Witten, Adv. Theor. Math. Phys. \textbf{2}, 253 (1998)

\bibitem{Cho:2002hq}
Y.M. Cho, I.P. Neupane, Phys. Rev. \textbf{D66}, 024044 (2002).
\newblock \doi{10.1103/PhysRevD.66.024044}

\bibitem{Neupane:2009zz}
I.P. Neupane, N.~Dadhich, Class. Quantum Grav. \textbf{26}, 015013 (2009).
\newblock \doi{10.1088/0264-9381/26/1/015013}

\bibitem{Kodama:2003jz}
H.~Kodama, A.~Ishibashi, Prog. Theor. Phys. \textbf{110}, 701 (2003).
\newblock \doi{10.1143/PTP.110.701}

\bibitem{Ishibashi:2003ap}
A.~Ishibashi, H.~Kodama, Prog. Theor. Phys. \textbf{110}, 901 (2003).
\newblock \doi{10.1143/PTP.110.901}

\bibitem{Gibbons:2002pq}
G.~Gibbons, S.A. Hartnoll, Phys. Rev. \textbf{D66}, 064024 (2002).
\newblock \doi{10.1103/PhysRevD.66.064024}

\bibitem{Policastro:2002se}
G.~Policastro, D.T. Son, A.O. Starinets, J. High Energy Phys. \textbf{09}, 043
  (2002).
\newblock \doi{10.1088/1126-6708/2002/09/043}

\bibitem{Rubin:1983be}
M.A. Rubin, C.R. Ord\'o\~{n}ez, J. Math. Phys. \textbf{25}, 2888 (1984).
\newblock \doi{10.1063/1.526034}

\bibitem{Rubin:1984tc}
M.A. Rubin, C.R. Ord\'o\~{n}ez, J. Math. Phys. \textbf{26}, 65 (1985).
\newblock \doi{10.1063/1.526749}

\bibitem{Higuchi:1986wu}
A.~Higuchi, J. Math. Phys. \textbf{28}, 1553 (1987).
\newblock \doi{10.1063/1.527513}.
\newblock (Erratum: ibid.\ {\bf 43}, 6385 (2002))

\bibitem{Myers:1988ze}
R.C. Myers, J.Z. Simon, Phys. Rev. \textbf{D38}, 2434 (1988).
\newblock \doi{10.1103/PhysRevD.38.2434}

\bibitem{Cai:2001dz}
R.G. Cai, Phys. Rev. \textbf{D65}, 084014 (2002).
\newblock \doi{10.1103/PhysRevD.65.084014}

\bibitem{Cai:2003kt}
R.G. Cai, Phys. Lett. \textbf{B582}, 237 (2004).
\newblock \doi{10.1016/j.physletb.2004.01.015}

\bibitem{Ge:2008ni}
X.H. Ge, Y.~Matsuo, F.W. Shu, S.J. Sin, T.~Tsukioka, J. High Energy Phys.
  \textbf{10}, 009 (2008).
\newblock \doi{10.1088/1126-6708/2008/10/009}

\bibitem{Kubiznak:2012wp}
D.~Kubiz\v{n}\'ak, R.B. Mann, J. High Energy Phys. \textbf{07}, 033 (2012).
\newblock \doi{10.1007/JHEP07(2012)033}

\bibitem{Kubiznak:2016qmn}
D.~Kubiz\v{n}\'ak, R.B. Mann, M.~Teo, Class. Quantum Grav. \textbf{34}, 063001
  (2017).
\newblock \doi{10.1088/1361-6382/aa5c69}

\bibitem{Hendi:2017fxp}
S.H. Hendi, R.B. Mann, S.~Panahiyan, B.~Eslam~Panah, Phys. Rev. \textbf{D95},
  021501 (2017).
\newblock \doi{10.1103/PhysRevD.95.021501}

\bibitem{Hawking:1982dh}
S.W. Hawking, D.N. Page, Commun. Math. Phys. \textbf{87}, 577 (1983).
\newblock \doi{10.1007/BF01208266}

\bibitem{Birmingham:1998nr}
D.~Birmingham, Class. Quantum Grav. \textbf{16}, 1197 (1999).
\newblock \doi{10.1088/0264-9381/16/4/009}

\bibitem{Konoplya:2017ymp}
R.A. Konoplya, A.~Zhidenko, Phys. Rev. \textbf{D95}, 104005 (2017).
\newblock \doi{10.1103/PhysRevD.95.104005}

\bibitem{Hu:2013cia}
C.~Hu, X.~Zeng, X.~Liu, Sci. China Phys. Mech. Astron. \textbf{56}, 1652
  (2013).
\newblock \doi{10.1007/s11433-013-5107-4}

\bibitem{Zeng:2016aly}
S.~He, L.F. Li, X.X. Zeng, Nucl. Phys. \textbf{B915}, 243 (2017).
\newblock \doi{10.1016/j.nuclphysb.2016.12.005}

\bibitem{Frassino:2014pha}
A.M. Frassino, D.~Kubiz\v{n}\'ak, R.B. Mann, F.~Simovic, J. High Energy Phys.
  \textbf{09}, 080 (2014).
\newblock \doi{10.1007/JHEP09(2014)080}

\bibitem{Dotti:2004sh}
G.~Dotti, R.J. Gleiser, Class. Quantum Grav. \textbf{22}, L1 (2005).
\newblock \doi{10.1088/0264-9381/22/1/L01}

\bibitem{Dotti:2005sq}
G.~Dotti, R.J. Gleiser, Phys. Rev. \textbf{D72}, 044018 (2005).
\newblock \doi{10.1103/PhysRevD.72.044018}

\bibitem{Knittel:1977}
D.R. Knittel, S.P. Pack, S.H. Lin, L.~Eyring, J. Chem. Phys. \textbf{67}, 134
  (1977).
\newblock \doi{10.1063/1.434557}

\bibitem{Bertotti:1998}
G.~Bertotti, \emph{{Hysteresis in Magnetism: For Physicists, Materials
  Scientists, and Engineers}} (Academic Press, San Diego, CA, USA, 1998)

\bibitem{Nguyen:2008}
C.T. Nguyen, F.~Desgranges, N.~Galanis, G.~Roy, T.~Mar\'e, S.~Boucher,
  H.~Angue~Mintsa, Int. J. Therm. Sci. \textbf{47}, 103 (2008).
\newblock \doi{10.1016/j.ijthermalsci.2007.01.033}

\end{thebibliography}

\end{document}